\pdfoutput=1
\documentclass[11pt, letterpaper]{article}

\usepackage[margin=1in]{geometry}
\usepackage[T1]{fontenc}
\usepackage{lmodern}
\usepackage{amsmath,amssymb,amsthm}
\usepackage[sort,numbers]{natbib}
\usepackage{authblk}
\usepackage[ruled,linesnumbered]{algorithm2e}
\usepackage{hyperref}
\hypersetup{pdftitle={Deterministic Root-Finding over Finite Fields with Elliptic Curves}, pdfauthor={Pierre-Jean Spaenlehauer}}
\usepackage{tikz}
\usepackage{pgfplots}
\usepackage{stmaryrd}
\usepackage{xurl}

\mathchardef\UrlBigBreakPenalty=0
\mathchardef\UrlBreakPenalty=5000
\usepackage{enumitem}
\usepackage{booktabs}
\usepackage{array}
\SetEnumitemKey{algo}{noitemsep,nosep,rightmargin=\algomargin}
\SetKwInput{INPUT}{\hspace*{-2.5mm}Input}
\SetKwInput{OUTPUT}{\hspace*{-2.5mm}Output}

\pgfplotsset{width=7cm,compat=1.18}

\numberwithin{equation}{section}
\newtheorem{theorem}{Theorem}
\numberwithin{theorem}{section}
\newtheorem{proposition}[theorem]{Proposition}
\newtheorem{example}[theorem]{Example}
\newtheorem{lemma}[theorem]{Lemma}
\newtheorem{corollary}[theorem]{Corollary}
\newtheorem{definition}[theorem]{Definition}
\newtheorem{remark}[theorem]{Remark}
\newtheorem{conjecture}[theorem]{Conjecture}
\newtheorem{problem}[theorem]{Problem}

\newcommand*{\R}{\mathbb{R}}
\newcommand*{\Z}{\mathbb{Z}}
\newcommand*{\Q}{\mathbb{Q}}
\newcommand*{\F}{\mathbb{F}}

\newcommand*{\Otilde}{\widetilde O}

\newcommand{\intrange}[2]{\llbracket #1, #2\rrbracket}

\newcommand*{\FiniteFieldEmbedding}{\textup{\textsc{FiniteFieldEmbedding}}}
\newcommand*{\CyclotomicPrimaryFacto}{\textup{\textsc{CyclotomicPrimaryFacto}}}
\newcommand*{\TwoadicFactoWithEval}{\textup{\textsc{TwoadicFactoWithEval}}}
\newcommand*{\SplitDivisorTranslation}{\textup{\textsc{ECTranslation}}}
\newcommand*{\EllipticCurvePrimaryFactorization}{\textup{\textsc{ECPrimaryFacto}}}
\newcommand*{\EllipticCurvePrimaryFactorizationAtLevelm}{\textup{\textsc{ECPrimaryFactoAtLevelm}}}
\newcommand*{\EllipticCurveRootFinding}{\textup{\textsc{ECRootFinding}}}

\DeclareMathOperator{\GCD}{GCD}
\DeclareMathOperator{\LCM}{LCM}
\DeclareMathOperator{\val}{val}
\DeclareMathOperator{\order}{order}
\DeclareMathOperator{\exponent}{exponent}
\DeclareMathOperator{\odd}{odd}
\DeclareMathOperator{\fac}{fac}

\title{Deterministic Root-Finding over Finite Fields with Elliptic Curves}

\author{Pierre-Jean Spaenlehauer}
\affil{Université de Lorraine, CNRS, Inria}

\date{}

\begin{document}

\pagenumbering{roman}
\maketitle
\thispagestyle{empty}

\begin{abstract}
Finding the roots of a polynomial over a finite field $\mathbb F_q$ in deterministic polynomial time is a long-standing open problem. Classical deterministic methods work in the multiplicative group $\mathbb F_q^\times$ and are efficient only when $q-1$ is smooth, which is rarely the case.

We propose a method that replaces $\mathbb F_q^\times$ by the groups of rational points of an elliptic curve $E$ and of its quadratic twist $\widetilde E$. The computation splits into an offline phase, which depends only on $\mathbb F_q$ and outputs a short \emph{elliptic preprocessing datum} $\mathcal D$, and an online phase, which processes the input polynomial. Writing $\Xi(\mathcal D)$ for the largest prime factor of $\lvert E(\mathbb F_q)\rvert\cdot\lvert\widetilde E(\mathbb F_q)\rvert$, the online phase is deterministic and finds the roots of any $f\in\mathbb F_q[X]$ of degree $n$ within $\widetilde O(\Xi(\mathcal D)^2\,n\log^2 q+n^\omega\log q+n\log^3 q)$ operations in $\mathbb F_q$. A precursor of our method is an algorithm by Rónyai, which recovers the roots through Galois computations over number fields, relies on GRH, and comes with a polynomial bound of unspecified degree; in contrast, our online phase works entirely over $\mathbb F_q$ and is unconditional. Moreover, $\mathcal D$ can be certified deterministically and unconditionally in time polynomial in $\log q$ and $\Xi(\mathcal D)$.

We conjecture that every $\mathbb F_q$ of large enough characteristic admits a datum with $\Xi(\mathcal D)\leq(\log q)^{O(1)}$, and support this with a heuristic and with experiments. Constructing such a datum in polynomial time remains open.
\end{abstract}

\clearpage
\tableofcontents
\thispagestyle{empty}
\clearpage
\pagenumbering{arabic}

\section{Introduction}

\subsection{Problem statement}

\paragraph{Problem statement.} Let $q = p^e$ be a prime power with $p > 3$ (for $p \leq 3$, Berlekamp's algorithm~\cite{berlekamp1970factoring} is already deterministic and polynomial-time), and $\F_q$ denote the finite field with $q$ elements. A univariate polynomial $f(X) \in \F_q[X]$ is \emph{separable} if it is coprime with its derivative, and \emph{split} if all its irreducible factors have degree $1$.

In this paper, we study deterministic algorithms for the following problem:

\begin{problem}[Root-finding over $\F_q$]\label{prob:rootfinding}
Given a prime power $q$ and a split, separable polynomial $f(X) \in \F_q[X]$, compute the set of roots of $f$ in $\F_q$, i.e., the set of $x \in \F_q$ such that $f(x) = 0$.
\end{problem}

This problem lies at the core of polynomial factorization over $\F_q$: indeed, factorization can be reduced to root-finding in deterministic polynomial time via squarefree factorization and Berlekamp's subalgebra (see~\cite{berlekamp1970factoring} and~\cite[Sec.~14]{von2003modern}).

\medskip

\paragraph{Motivation.} Root-finding over finite fields is a central problem in algebraic computation. Although efficient randomized algorithms have been known for decades, no deterministic polynomial-time algorithm has been found (see however the recent claim discussed at the end of Section~\ref{sec:intro_priorwork}). Finding one would constitute a step toward the broader goal of derandomizing algebraic algorithms.
Root-finding and polynomial factorization over finite fields are subroutines of many algorithms in computational algebra. The absence of a deterministic polynomial-time solution propagates to higher-level algorithms.
A telling example is the algorithmic structure theory of finite-dimensional associative algebras over $\F_q$: finding zero divisors, decomposing a semisimple algebra into simple components, exhibiting an explicit isomorphism between a simple algebra and a full matrix algebra, or computing a common invariant subspace of a family of linear maps are all solved by \emph{$f$-algorithms}, i.e.\ deterministic polynomial-time algorithms with access to an oracle for factoring univariate polynomials over finite fields~\cite{friedl1985polynomial, ronyai1990computing, eberly1991decomposition}. Factorization is itself a special case of the zero-divisor problem~\cite{ronyai1987simple}. These problems are therefore exactly as hard as Problem~\ref{prob:rootfinding}.

Deterministic algorithms can also be used in cryptography, where $\F_q$ is sometimes fixed by a standard, so that a field-dependent precomputation can be performed. For instance, constant-time hashing into elliptic curves extracts square roots by a deterministic variant of the Tonelli--Shanks algorithm, which relies on a nonsquare of $\F_q$ fixed in advance~\cite[App.~I.4]{rfc9380}; our preprocessing datum plays a similar role for polynomials of arbitrary degree.

\medskip

\paragraph{Context.} Polynomial factorization and root-finding over finite fields have been studied for several decades. Berlekamp is often credited with initiating the modern algorithmic study of these problems~\cite{berlekamp1967factoring}, although some key ideas already appear in the works of Gauss and Legendre. We refer to~\cite{von2001factoring} for a historical survey of the area.
Over the years, fast probabilistic algorithms have been developed, whereas deterministic polynomial-time algorithms are only known in restricted settings, often under GRH. Such algorithms exist when the characteristic of $\F_q$ is fixed~\cite{berlekamp1970factoring, shoup1991fast}, when $f(X)$ has a bounded number of irreducible factors~\cite{RONYAI1988391}, or when $f(X)$ arises as the reduction of an integer polynomial with suitable Galois properties~\cite{ronyai1992galois}, the last two under GRH. Without GRH, polynomials over a prime field $\F_p$ can be factored deterministically in time polynomial in $\deg(f)$ and $p^{1/2}$~\cite{shoup1990deterministic}, which is exponential in $\log p$.
For general finite fields, the best known deterministic algorithm is that of~\cite{evdokimov1994factorization}, which runs under GRH in time polynomial in $\log q$ and $\deg(f)^{\log\deg(f)}$. Ivanyos, Karpinski and Saxena interpret the combinatorics underlying this algorithm in terms of \emph{$m$-schemes}~\cite{ivanyos2009schemes}, which yield a deterministic polynomial-time algorithm under GRH for polynomials of prime degree $n$ such that $n-1$ is smooth; this was extended in~\cite{arora2014deterministic} to the case where $n-1$ has a large $r$-smooth divisor, in time $(n^r\log q)^{O(1)}$. Ivanyos, Karpinski, Rónyai and Saxena eliminate GRH from most of these results~\cite{ivanyos2012trading}, at the price of a weaker output: for instance, in deterministic time polynomial in $\log q$ and $\deg(f)^{\log\deg(f)}$, their algorithm returns either a nontrivial factor of $f(X)$ or a nontrivial automorphism of $\F_q[X]/(f(X))$. Another approach is given by Camion's algorithm~\cite{CAMION1983149, camion1983improving}, which provides a deterministic variant of the Cantor--Zassenhaus algorithm, but relies on a precomputed subset of $\F_q$ for which no algorithm polynomial in $\log q$ is known.

\subsection{Main contributions}\label{sec:intro_contrib}

Our main contribution is a deterministic root-finding algorithm for $\F_q$ that splits into two phases: an \emph{offline} phase, which depends only on the field, and an \emph{online} phase, which takes the polynomial $f(X)$ as input. The interface between the two is the following notion.

\begin{definition}[Elliptic preprocessing datum]\label{def:datum}
An elliptic preprocessing datum for $\F_q$ is a tuple
\[
\mathcal D=(\beta,E,\widetilde E, N,\widetilde N,
\fac(N),\fac(\widetilde N),
G_1,G_2,\widetilde G_1,\widetilde G_2),
\]
where $\beta\in\F_q^\times$ is a nonsquare, $E$ is an elliptic curve given by a short Weierstrass equation, $\widetilde E$ is its quadratic twist w.r.t. $\beta$; the integers $N=\lvert E(\F_q)\rvert$ and
$\widetilde N=\lvert\widetilde E(\F_q)\rvert$ are the orders of the groups of rational points, with factorizations $\fac(N)$ and $\fac(\widetilde N)$;
finally, $G_1, G_2\in E(\F_q)$ and
$\widetilde G_1,\widetilde G_2\in \widetilde E(\F_q)$ are bases for their respective groups, i.e.
$E(\F_q)=\langle G_1\rangle\oplus\langle G_2\rangle$ and
$\widetilde E(\F_q)=\langle\widetilde G_1\rangle\oplus\langle\widetilde G_2\rangle$. We allow $G_2=\infty$, so that the cyclic case is covered by letting $G_1$ be a generator.
\end{definition}

We let $\Xi(\mathcal D)$ denote the largest prime dividing $N\cdot\widetilde N$. The cost of the online phase depends on the field essentially through this quantity.

Our main result is the following.

\begin{theorem}[Elliptic curve root-finding, online phase]
\label{thm:intro-rootfinding}
There is a deterministic algorithm (Algorithm~\ref{algo:ellipticcurverootfinding}) which, given a split separable
polynomial $f\in\F_q[X]$ and an elliptic preprocessing datum
$\mathcal D$ for $\F_q$, outputs all roots of $f$ in $\F_q$.

Excluding the cost of constructing and verifying $\mathcal D$, the
algorithm performs at most
\[
\widetilde O\!\left(
\Xi(\mathcal D)^2\deg(f)\log^2 q
+\deg(f)^\omega\log q
+\deg(f)\log^3 q
\right)
\]
arithmetic operations in $\F_q$.
\end{theorem}

The online phase is unconditional: it relies on no unproven hypothesis, and in particular not on GRH. This contrasts with the elliptic-curve method in~\cite{ronyai1992galois}, which relies on GRH through Galois-theoretic machinery over number fields. Moreover, apart from factors polynomial in $\log q$, the bound depends on the field only through $\Xi(\mathcal D)$, so that a datum with small $\Xi(\mathcal D)$ yields an efficient online phase regardless of the divisibility properties of $q-1$.

The difficulty is thus displaced to the offline phase, which we address from the angles of verification, existence, and construction.

\paragraph{Verification.} Preprocessing data are efficiently verifiable: their validity can be checked deterministically and unconditionally in time polynomial in $\log q$ and $\Xi(\mathcal D)$ (Proposition~\ref{prop:verification}).

This distinguishes our approach from Camion's framework~\cite{CAMION1983149, camion1983improving}, another deterministic method whose precomputation depends only on $\F_q$. There, the precomputed object is a subset of $\F_q$ required to separate the roots of every input polynomial; we are not aware of any procedure certifying it in time polynomial in $\log q$, so that the precomputation cannot be decoupled from its verification.

\paragraph{Existence.} The existence of good preprocessing data (i.e. with small $\Xi(\mathcal D)$) is related to the existence of pairs of smooth orders in the Hasse--Weil interval, which contains the order of $E(\F_q)$ for every elliptic curve $E$ over $\F_q$ (Hasse's theorem, see Section~\ref{sec:prelim}):

\begin{proposition}\label{prop:datum_existence}
Let $(N, \widetilde N)$ be a pair of integers in the Hasse--Weil interval $[q+1-2\sqrt q,\, q+1+2\sqrt q]$ such that $N+\widetilde N = 2(q+1)$ and $N\not\equiv 1\bmod p$. Then there exists an elliptic preprocessing datum $\mathcal D$ for $\F_q$ with $\lvert E(\F_q)\rvert = N$ and $\lvert\widetilde E(\F_q)\rvert = \widetilde N$.
\end{proposition}

\begin{proof}
Set $U := q+1-N$. That $N$ lies in the Hasse--Weil interval means $\lvert U\rvert \leq 2\sqrt q$, and $N\not\equiv 1 \bmod p$ means that $U$ is coprime to $q$. By~\cite[Thm.~4.1]{waterhouse1969abelian}, there exists an elliptic curve $E$ over $\F_q$ with $\lvert E(\F_q)\rvert = N$, and since $p>3$ it admits a short Weierstrass model. Pick any nonsquare $\beta\in\F_q^\times$, and let $\widetilde E$ be the quadratic twist of $E$ with respect to $\beta$. The Frobenius trace of $E$ is $U$, hence that of $\widetilde E$ is $-U$ (Section~\ref{sec:prelim}), whence $\lvert\widetilde E(\F_q)\rvert = q+1+U = 2(q+1)-N = \widetilde N$. Finally, $E(\F_q)$ and $\widetilde E(\F_q)$ are finite abelian groups of rank at most $2$, so each admits a basis, with the convention that $G_2 = \infty$ in the cyclic case.
\end{proof}

Proposition~\ref{prop:datum_existence} turns the existence of a good preprocessing datum into a purely number-theoretic question: does the Hasse--Weil interval contain pairs of smooth integers summing to $2(q+1)$? We conjecture that it does.

\begin{conjecture}\label{conj:main_conj}
  There exist constants $c>0$ and $p_0$ such that, for every prime $p\geq p_0$ and every integer $e\geq 1$, letting $q=p^e$, there exists an integer $U\in[0, 2\sqrt q]$ coprime to $q$ such that both $q+1-U$ and $q+1+U$ are $(\log q)^{c}$-smooth.
\end{conjecture}

Section~\ref{sec:distrib_smooth} identifies this statement as a constrained instance of a question of Erd\H os and Graham on sums of two smooth integers, and discusses what the standard heuristics predict for the exponent $c$; Section~\ref{sec:expe_conj} reports experiments. Combined with Proposition~\ref{prop:datum_existence}, it gives:

\begin{corollary}\label{cor:datum_exists}
Under Conjecture~\ref{conj:main_conj}, every finite field $\F_q$ of sufficiently large characteristic admits an elliptic preprocessing datum $\mathcal D$ with $\Xi(\mathcal D)\leq(\log q)^{O(1)}$.
\end{corollary}

\paragraph{Construction.} There remains the task of producing such a datum deterministically. Section~\ref{sec:precomp} shows that, once a curve with smooth group orders is known, all the remaining components of the datum can be computed in deterministic polynomial time under GRH when $q$ is prime. Constructing the curve is therefore the obstruction:

\begin{problem}[Deterministic preprocessing]\label{prob:precomp}
Does there exist a deterministic polynomial-time algorithm that, on
input $q$, constructs an elliptic preprocessing datum $\mathcal D$
with $\Xi(\mathcal D)\leq(\log q)^{O(1)}$?
\end{problem}

\begin{corollary}
Under a positive answer to Problem~\ref{prob:precomp}, root-finding over $\F_q$ (Problem~\ref{prob:rootfinding}) is solvable deterministically in time polynomial in $\log q$ and $\deg(f)$.
\end{corollary}

Our main result can therefore be read as a reduction: deterministic root-finding over $\F_q$ reduces to the offline construction of an elliptic curve whose group of rational points and that of its quadratic twist both have smooth order.

\medskip

\paragraph{Implementation.} Our algorithmic description is sufficiently precise to enable a direct implementation in computer algebra systems. We provide a prototype implementation in Magma, available at~\cite{implementation}. Our experiments show that the algorithm is not restricted to special values of $q$, in contrast to cyclotomic methods, which rely on smoothness assumptions on $q-1$: for general prime powers, our unoptimized implementation performs deterministic root-finding computations in practice, and for instance handles prime fields of size larger than $2^{63}$ on a standard laptop.

\subsection{Comparison with prior work}\label{sec:intro_priorwork}

\paragraph{Cyclotomic methods.}
Cyclotomic methods reduce root-finding to computations in smooth subgroups of $\F_q^\times$, and their cost is governed by the largest prime factor of $q-1$. This line was initiated by~\cite{moenck1977efficiency}, for prime powers $q$ such that $q-1$ is divisible by a large power of $2$, and turned into a general framework by Mignotte and Schnorr~\cite{mignotte1988calcul} and by von zur Gathen~\cite{von1987factoring}. These algorithms use the decomposition
\[
\F_q^\times \simeq \Z/\ell_1^{m_1}\Z \times \cdots \times \Z/\ell_r^{m_r}\Z,
\]
and the fact that, writing $\pi_i : \F_q^\times \to \Z/\ell_i^{m_i}\Z$ for the projection onto the $i$-th factor, the polynomial whose roots are exactly the $x \in \F_q^\times$ with $\pi_i(x) = a$ is a \emph{binomial}
\[
X^{(q-1)/\ell_i^{m_i}} - \gamma,
\]
so that its GCD with $f(X)$ is obtained from a single binary exponentiation. The construction was extended in~\cite{bach2001factoring} to fields for which the value at $q$ of some cyclotomic polynomial has no large prime factor.

A parallel line of work is based on (generalized) Graeffe transforms: replacing $f(X)$ by the polynomial whose roots are the $\ell$-th powers of its roots pushes the roots into a subgroup of index $\ell$, and iterating over the prime divisors of $q-1$ produces a sequence ending at $(X-1)^{\deg(f)}$. The difficulty is to trace the roots back through this sequence, which~\cite{grenet2015randomized, grenet2016deterministic, van2021computing} achieve using the \emph{tangent Graeffe transform}; a related algorithm is given in~\cite[Sec.~3.2]{davenport2016generalised}, whose backward step calls an external root-finding procedure. These methods require $q-1$ to be smooth.

Our algorithm uses the recursive framework of~\cite{mignotte1988calcul} and changes the group it acts upon. Section~\ref{sec:ov_dictionary} describes the correspondence, and Appendix~\ref{sec:app_MS} restates the cyclotomic algorithm in the formalism used here.

\paragraph{Replacing cyclotomy by elliptic curves.}
Substituting an elliptic curve of smooth group order for a smooth cyclotomic group is a classical technique in computational number theory. The best-known instance is the elliptic curve method for integer factorization~\cite{ecm}, which replaces the smoothness assumption on $p-1$ in Pollard's algorithm by one on $\lvert E(\F_p)\rvert$; another is elliptic curve primality proving~\cite{goldwasser1986almost, atkin1993elliptic}, which replaces the Pocklington test (and the partial factorization of $p-1$ it requires) by a condition on a suitably chosen curve.

Beyond these two examples, an elliptic-curve version of the AKS primality test was designed in~\cite{ezome2013elliptic}. A more recent instance arises in FFT-based multiplication: multiplying in $\F_q[X]$ by the classical FFT requires $\F_q^\times$ to contain many elements of smooth order, and when this fails one must build ``synthetic'' roots of unity~\cite{schonhage1971schnelle, furer2007faster, harvey2021integer}. The works of~\cite{ECFFT1} and~\cite{couveignes2026elliptic} show that elliptic curves with many rational points of smooth order can be used instead. In all of these, as here, the arithmetic of $q-1$ is traded for the arithmetic of a group of rational points on an elliptic curve.

The closest precursor to the present work is the algorithm sketched at the end of~\cite{ronyai1992galois} for deterministic root-finding over $\F_p$. It may be read as an elliptic analogue of the Graeffe transform: given $f(X)$, a curve $E$, and a prime $\ell$ dividing $\lvert E(\F_{p^2})\rvert$, it computes the polynomial whose roots are the $x$-coordinates of the points $[\ell\cdot(x,y)]$, where $x$ ranges over the roots of $f$ and $y^2 = x^3+ax+b$. Iterating projects the roots into subgroups of decreasing size, and recovering them amounts to computing points $P$ with $[\ell\cdot P] = Q$. The resulting complexity is polynomial in $\log p$, $\deg(f)$ and the largest prime factor of $\lvert E(\F_{p^2})\rvert$, although the exponents are not made explicit.

Our algorithm differs from this one in two respects. First, its root-recovery step proceeds by lifting the problem to characteristic zero and exploiting the Galois structure of the torsion of a lift of $E$ over $\Q$~\cite[Cor.~5.6]{ronyai1992galois}; this requires computing Galois groups of division polynomials, which is polynomial-time (the $\ell$-torsion lies in an extension of degree less than $\ell^4$) but prohibitively expensive in practice, and is only known to be effective under GRH. Our algorithm performs all the computations over $\F_q$: it uses only the intrinsic group structure of elliptic curves over finite fields, and its online phase is unconditional. Second, we track the complexity of every step, which yields the explicit bound of Theorem~\ref{thm:intro-rootfinding}, and the resulting algorithm is simple enough to be implemented.

Variants using isogenies in place of multiplication maps were proposed in~\cite[Sec.~3.3]{davenport2016generalised}. There the roots are embedded on $E$ using Icart's map~\cite{icart2009hash}, which forces $p \equiv 2 \bmod 3$. Inverting the isogeny, however, requires an external root-finding procedure.

\paragraph{Encoding field elements as points.}
The lifting of Section~\ref{sec:ov_lifting} is a deterministic encoding of $\F_q$ into the rational points of a curve and its twist. Such encodings have been considered in cryptography, where hashing into elliptic curves is a well-studied primitive: see~\cite{shallue2006construction} and Icart's map~\cite{icart2009hash}.

Compared with Icart's map, our method imposes no congruence condition on $p$, and it is injective, since $x$ is recovered from the $x$-coordinate of its image; moreover, up to the sign of $y$, it reaches every affine point of $E(\F_q)$ and of $\widetilde E(\F_q)$, except the (at most three) points of order $2$ of $\widetilde E(\F_q)$. Nevertheless, it is not a substitute for a hash into a prescribed curve.

\paragraph{Other arithmetico-geometric approaches.}
Root-finding and factorization have also been approached through other objects of arithmetic geometry. Kayal proposed a factoring algorithm based on point-counting on elliptic curves, described in~\cite{wolfsdorf2017factorising}, which Poonen extended to abelian varieties~\cite{poonen2017using}. A separate line is based on Drinfeld modules: it originates in~\cite{panchishkin1989algorithm}, was rediscovered in~\cite{van2004factoring}, and modern fast probabilistic algorithms in this framework are given in~\cite{narayanan2018polynomial, narayanan2019polynomial, doliskani2021drinfeld}. Root-finding also arises when computing isomorphisms between representations of finite fields: constructing $\F_q[X]/P(X)\simeq\F_q[X]/Q(X)$ amounts to finding a root of $Q$ in $\F_q[X]/P(X)$. Elliptic-curve methods for this problem are described in~\cite{Pinch92c}. Finally, one of the first applications of Schoof's algorithm~\cite{schoof1985elliptic} was the computation of square roots modulo a prime, that is, root-finding for $X^2-a$, the smallest nontrivial instance of Problem~\ref{prob:rootfinding}.

\paragraph{A recent claim by OpenAI.}
A recent preprint by OpenAI~\cite{openai2026factorization} claims an unconditional deterministic algorithm that factors a polynomial $f\in\F_p[X]$ in $O\bigl(((\deg(f)+1)\lceil\log_2 p\rceil)^{10^{12}}\bigr)$ bit operations. Its unconditionality rests on a uniform zero-free region for Hecke $L$-functions established in a companion preprint, and its algebraic part alone is claimed to yield a polynomial-time algorithm under GRH. At the time of writing, this claim has not been reviewed or validated by the community, and no formalization of its proof is available.

\section{Technical overview}\label{sec:overview}

This section is a self-contained account of the paper. We follow the algorithm from end to end, state the intermediate results with their complexities, and show how they assemble into Theorem~\ref{thm:intro-rootfinding}; we then turn to the offline phase. Proofs are deferred to Sections~\ref{sec:lifting}--\ref{sec:precomp}.

Throughout, $f(X)\in\F_q[X]$ is the split separable input polynomial and $n := \deg(f)$. We fix an elliptic preprocessing datum $\mathcal D$ (Definition~\ref{def:datum}) with curve $E : Y^2 = h(X)$, where $h(X) = X^3+aX+b$, quadratic twist $\widetilde E$, group orders $N$ and $\widetilde N$, nonsquare $\beta\in\F_q$, and we abbreviate $\Xi := \Xi(\mathcal D)$.

\begin{figure}[t]
\centering
\fbox{\begin{minipage}{0.95\linewidth}
\textbf{Background on elliptic curves.} Let $E$ be the elliptic curve defined by the equation ${Y^2 = h(X)}$, where $h(X) = X^3+aX+b$ with $a,b\in\F_q$ and $4a^3+27b^2\neq 0$, and let $\beta\in\F_q$ be a nonsquare. The facts below suffice to read this section; Section~\ref{sec:prelim} gives details and references.
\begin{itemize}[leftmargin=*, itemsep=2pt, topsep=4pt]
\item \emph{Group law.} The set $E(\F_q)$ consists of the solutions $(x,y)\in\F_q^2$ of $y^2 = h(x)$, together with a point at infinity $\infty$. It is an abelian group with identity $\infty$, in which the opposite of $(x,y)$ is $(x,-y)$; the points of order $2$ are thus the $(x,0)$ with $h(x)=0$. The coordinates of $[P+Q]$ are rational functions of those of $P$ and $Q$ (see Section~\ref{sec:ov_translation}), so adding two points takes $O(1)$ operations in $\F_q$.
\item \emph{Group structure.} $E(\F_q)\simeq\Z/m_1\Z\times\Z/m_2\Z$ with $m_2\mid m_1$: unlike $\F_q^\times$, this group need not be cyclic, but it is generated by two points, such as the points $G_1, G_2$ of the datum.
\item \emph{Group order.} By Hasse's theorem, the order of $E(\F_q)$ is $N = q+1-U$, where the integer $U$, called the trace of $E$, satisfies $\lvert U\rvert\leq 2\sqrt q$. Thus $N$ lies in the Hasse--Weil interval $[q+1-2\sqrt q,\, q+1+2\sqrt q]$.
\item \emph{Quadratic twist.} $\widetilde E$ is the curve $Y^2 = X^3+a\beta^{-2}X+b\beta^{-3}$, whose right-hand side is $\beta^{-3}h(\beta X)$. Every $x\in\F_q$ with $h(x)\neq 0$ is the abscissa of two points of $E$ if $h(x)$ is a square, and $x/\beta$ is the abscissa of two points of $\widetilde E$ otherwise; each root $x$ of $h$ gives the point $(x,0)$ of $E$ and the point $(x/\beta,0)$ of $\widetilde E$. Hence the order $\widetilde N$ of $\widetilde E(\F_q)$ satisfies $N+\widetilde N = 2(q+1)$, that is, $\widetilde N = q+1+U$.
\item \emph{Division polynomials.} For $r\geq 1$, the division polynomial $\psi_r(X,Y)$ vanishes at a point $P\neq\infty$ of $E$ if and only if $[rP]=\infty$. It plays the role of $X^r-1$ for $\F_q^\times$, and its value at a point can be computed with $O(\log r)$ ring operations by recurrence relations, as $X^r$ is by binary exponentiation.
\end{itemize}
\end{minipage}}
\end{figure}

\subsection{From cyclotomy to elliptic curves}\label{sec:ov_dictionary}

The deterministic algorithm of Mignotte and Schnorr~\cite{mignotte1988calcul} separates the roots of $f$ according to their images in the successive quotients $\F_q^\times[\ell^i]/\F_q^\times[\ell^{i-1}]$ of the $\ell$-Sylow subgroup of $\F_q^\times$, for each prime $\ell$ dividing $q-1$. The group $\F_q^\times$ is \emph{cyclic}, so each such quotient has exactly $\ell$ classes; and, as recalled in Section~\ref{sec:intro_priorwork}, the polynomial cutting out one class is a binomial, whose reduction modulo $f(X)$ costs one binary exponentiation. However, the recursion runs over the prime divisors of $q-1$, so the cost is governed by the largest of them.

We keep the recursion and replace $\F_q^\times$ by groups of rational points on elliptic curves. Every $x\in\F_q$ with $h(x)\neq 0$ lifts to a rational point of $E$ or of $\widetilde E$, according to whether $h(x)$ is a square, so the pair $E(\F_q), \widetilde E(\F_q)$ takes over the role of $\F_q^\times$, and the relevant smoothness parameter becomes $\Xi$. Since the pairs of group orders realized by a curve and its twist range over almost the whole Hasse--Weil interval (Proposition~\ref{prop:datum_existence}), this is a condition we may hope to satisfy for every $q$, whereas $q-1$ is rigidly fixed by $q$.

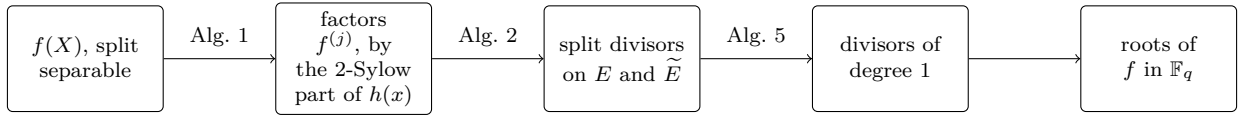
\begin{figure}[ht]
\centering
\begin{tikzpicture}[x=1cm, y=1cm]
  \tikzset{
    box/.style={draw, rounded corners=2pt, align=center, inner sep=3pt,
                text width=1.85cm, minimum height=1.4cm, font=\scriptsize},
    lbl/.style={font=\scriptsize, midway, above=1pt}
  }
  \node[box] (a) at (0,0)     {$f(X)$, split separable};
  \node[box] (b) at (3.55,0)  {factors $f^{(j)}$, by the $2$-Sylow part of $h(x)$};
  \node[box] (c) at (7.1,0)   {split divisors on $E$ and $\widetilde E$};
  \node[box] (d) at (10.65,0) {divisors of degree $1$};
  \node[box] (e) at (14.2,0)  {roots of $f$ in $\F_q$};
  \draw[->] (a) -- node[lbl] {Alg.~\ref{algo:TwoadicFactoWithEval}} (b);
  \draw[->] (b) -- node[lbl] {Alg.~\ref{algo:curveembedding}} (c);
  \draw[->] (c) -- node[lbl] {Alg.~\ref{algo:ellipticcurveprimaryfacto}} (d);
  \draw[->] (d) -- node[lbl] {} (e);
\end{tikzpicture}
\caption{The online phase. The first two steps are the subject of Section~\ref{sec:ov_lifting}, the third of Sections~\ref{sec:ov_translation} and~\ref{sec:ov_recursion}.}
\label{fig:pipeline}
\end{figure}

Table~\ref{tab:dictionary} lists the substitutions this requires, and Figure~\ref{fig:pipeline} shows a bird's-eye view of the resulting algorithm. Two subroutines are more delicate than their multiplicative counterparts: translating by a group element (Section~\ref{sec:ov_translation}), and the torsion test (Section~\ref{sec:ov_recursion}). 

\begin{table}[ht]
\centering
\renewcommand{\arraystretch}{1.25}
\begin{tabular}{@{}>{\raggedright\arraybackslash}p{0.45\textwidth}>{\raggedright\arraybackslash}p{0.49\textwidth}@{}}
\toprule
\textbf{Cyclotomic (Mignotte--Schnorr)} & \textbf{Elliptic (this paper)} \\
\midrule
$\F_q^\times$, cyclic of order $q-1$
  & $E(\F_q)$ and $\widetilde E(\F_q)$, of rank $\leq 2$ and orders $N,\widetilde N$ \\
Efficient when $q-1$ is smooth
  & Efficient when $N\widetilde N$ is smooth, i.e.\ when $\Xi$ is small \\
A root $x$ is an element of $\F_q^\times$
  & A root $x$ is lifted to a point $(x,v(x))$ of $E(\F_q)$ or of $\widetilde E(\F_q)$ (Section~\ref{sec:ov_lifting}) \\
A set of roots is a split polynomial $u(X)$
  & A set of points in $E(\F_q)$ or $\widetilde E(\F_q)$ is a split divisor, encoded by a pair $(u(X),v(X))$ (Section~\ref{sec:ov_divisors}) \\
Translation is multiplication by $\gamma\in\F_q^\times$, a ring operation modulo $u(X)$
  & Translation by a point $Q$ applies the group law to generic points (Section~\ref{sec:ov_translation}) \\
Torsion test: gcd with $X^{(q-1)/\ell^{m}}-\gamma$, reduced by binary exponentiation
  & Torsion test: gcd with the division polynomial $\psi_{N/\ell^{m}}(X,Y)$, evaluated at a generic point \\
Each torsion quotient is cyclic: $\ell$ classes per level
  & The successive quotients of $E(\F_q)[\ell^\infty]$ have rank $\leq 2$: up to $\ell^2$ classes per level (Section~\ref{sec:ov_recursion}) \\
\bottomrule
\end{tabular}
\caption{The correspondence underlying our algorithm.}
\label{tab:dictionary}
\end{table}

We illustrate each step on a running example, which the reader may follow from
Example~\ref{ex:datum83} onwards. Magma code reproducing this example is provided in the file
\texttt{ExampleIntro.mgm} in~\cite{implementation}.

\begin{example}[Running example: the datum]\label{ex:datum83}
Take $q = 83$ and $h(X) = X^3+31X+21$, defining an elliptic curve $Y^2 = h(X)$ of Frobenius trace $U = 12$. Then
$N = q+1-U = 2^3\cdot 3^2$ and $\widetilde N = q+1+U = 2^5\cdot 3$,
so that $\Xi = 3$. We complete the datum with the nonsquare $\beta = 2$ and the bases $G_1 = (32,0)$, $G_2 = (59,71)$ of $E(\F_q)\simeq \Z/2\Z\times\Z/2^2\Z\times\Z/3^2\Z$. In the model $\widetilde h(X) = \beta^{-3}h(\beta X)$ used by Algorithm~\ref{algo:ellipticcurverootfinding}, the twist is $\widetilde E : Y^2 = X^3+70X+13$, with basis $\widetilde G_1 = (16,0)$ and $\widetilde G_2 = (6,20)$ of orders $2$ and $48$, so that $\widetilde E(\F_q)\simeq\Z/2\Z\times\Z/2^4\Z\times\Z/3\Z$. The input polynomial will be
\[
f(X) = X^{10} + 28X^9 + 32X^8 + 41X^7 + 78X^6 + 58X^5 + 4X^4 + 48X^3 + 7X^2 + 12X + 26.
\]
\end{example}

\subsection{Lifting the roots onto the curves}\label{sec:ov_lifting}

The common roots of $f$ and $h$ are the abscissas of the points of order $2$, read off the datum; we assume from now on that $f$ is coprime with $h$.

Lifting a root $x$ to a point of $E$ means producing a square root of $h(x)$; lifting all roots at once, without knowing any of them, means producing a single polynomial $v(X)$ with $v(X)^2 \equiv h(X) \bmod f(X)$, so that $x\mapsto (x, v(x))$ sends the roots of $f$ to points of $E(\F_q)$.

Such a $v$ exists as soon as $h(x)$ is a square for every root $x$ of $f$, by interpolation through arbitrary choices of square roots (Lemma~\ref{lem:separateEtwist}); if instead $h(x)$ is a nonsquare at every root, the same argument gives $\beta v(X)^2 \equiv h(X) \bmod f(X)$, and the roots lift to the twist. So we first split $f$ according to the quadratic character of $h$, and then make the interpolant effective.

\paragraph{Splitting using the quadratic character.}
Since $h(x)^{(q-1)/2} = \pm 1$ at every root $x$ of $f$, the two GCDs
\[
\GCD\bigl(h(X)^{(q-1)/2} \pm 1 \bmod f(X),\; f(X)\bigr)
\]
split $f$ according to whether $h(x)$ is a square, at the cost of one binary exponentiation and one GCD. Halving the exponent and recursing refines this into a factorization
\[
f(X) = \prod_{j\in S} f^{(j)}(X), \quad S\subset\{0,\ldots,2^{\val_2(q-1)}-1\},
\]
indexed by the exponent $j$ for which $h(x)\beta^{-j}$ has odd order (Algorithm~\ref{algo:TwoadicFactoWithEval}). This is exactly the Mignotte--Schnorr recursion for $\ell = 2$, applied to the values $h(x)$. It runs in $\Otilde\bigl(n\log^2 q \bigr)$ operations in $\F_q$; see Proposition~\ref{prop:twoadicfactowitheval}.

\paragraph{Interpolation.}
Once $j$ is known, Lemma~\ref{lem:TonelliShanks} gives a square root in closed form. If $\gamma$ is an inverse of $2$ modulo $(q-1)/2^{\val_2(q-1)}$ and $\order(z\beta^{-j})$ is odd, then
\[
y := (z\beta^{-j})^{\gamma}\cdot\beta^{\lfloor j/2\rfloor}
\]
is a square root of $z$ when $j$ is even, and of $z\beta^{-1}$ when $j$ is odd. This is a variant of Tonelli--Shanks in which the usual search is replaced by the known exponent $j$; being a formula, it can be applied to the generic point corresponding to the factor $f^{(j)}$: substituting $h(X)$ for $z$ and reducing modulo $f^{(j)}(X)$ produces the desired $v(X)$. Applying this to each factor and recombining by the Chinese Remainder Theorem yields Algorithm~\ref{algo:curveembedding}: it returns polynomials $u_0, u_1, v_0, v_1$ such that $f = u_0u_1$, the roots of $u_0$ lift to $E$ through $v_0$ and those of $u_1$ lift to $\widetilde E$ through $v_1$, in $\Otilde\bigl(n\log^2 q\bigr)$ operations in $\F_q$; see Proposition~\ref{prop:finitefieldembedding}.

\begin{example}[Running example: the lift]\label{ex:lift83}
Continuing Example~\ref{ex:datum83}, Algorithm~\ref{algo:TwoadicFactoWithEval} splits $f = f_E\cdot f_{\widetilde E}$, and Algorithm~\ref{algo:curveembedding} returns $v$ with $v(X)^2\equiv h(X)\bmod f_E(X)$, where
\[
f_E(X) = X^5 + 66X^4 + 77X^3 + 75X^2 + 29X + 35,
\qquad
v(X) = 38X^4 + 32X^3 + 29X^2 + 57X + 30,
\]
so that the five roots of $f_E$ are now points $(x,v(x))$ of $E(\F_{83})$.
\end{example}

\subsection{Representing sets of points}\label{sec:ov_divisors}

Everything that follows is phrased in terms of divisors; we summarize here the terminology of Definition~\ref{def:splitdivisor}. A \emph{split divisor} is a formal sum $D = \sum_i P_i$ of pairwise distinct points of $E(\F_q)$; we call it \emph{simple} if moreover it avoids $\infty$ and its points have pairwise distinct $x$-coordinates. A simple divisor is encoded by its \emph{$(u,v)$-coordinates}: the monic polynomial $u(X)$ whose roots are the $x$-coordinates of the $P_i$, and the interpolant $v(X)$ of degree less than $\deg(u)$ through their $y$-coordinates, which satisfies $v(X)^2\equiv h(X) \bmod u(X)$. A simple divisor of degree $d$ is thus given by two polynomials of degree at most $d$, and can be handled by polynomial arithmetic.

The restriction to simple divisors is needed to deal with the fact that two opposite points share an $x$-coordinate, so $u(X)$ cannot separate them. A general split divisor is therefore represented as a \emph{sum} of simple divisors with disjoint supports, plus a flag for $\infty$; our algorithms take and return such sums, and process each summand independently.

In this language, the problem reached at the end of Section~\ref{sec:ov_lifting} is Problem~\ref{prob:ECdivdecomp}: given $(u,v)$ with $u$ split separable and $v^2\equiv h \bmod u$, recover the corresponding points. It is the elliptic analogue of root-finding, and the object of Sections~\ref{sec:ov_translation} and~\ref{sec:ov_recursion}.

\subsection{Translating a split divisor}\label{sec:ov_translation}

In the multiplicative setting, translating a set of roots by $\gamma$ is achieved by replacing $u(X)$ by $\gamma^{\deg u}u(X/\gamma)$. We now describe how to perform the analogous operation on a curve. The $(u,v)$-coordinates of a simple divisor $D$ are a rational parametrization of its support: the points of $D$ are exactly the $(x,v(x))$, for $x$ ranging over the roots of $u(X)$. Translating $D$ by a rational point $Q = (x_Q,y_Q)$ therefore amounts to applying the group law to the \emph{generic} point $(X,v(X))$, all computations taking place in $\F_q[X]/u(X)$; a general split divisor is handled summand by summand. The chord-and-tangent formulas give
\[
s(X) = \frac{v(X)-y_Q}{X-x_Q},\qquad
\varphi_x(X) = s(X)^2 - X - x_Q,\qquad
\varphi_y(X) = -y_Q + s(X)\bigl(x_Q - \varphi_x(X)\bigr),
\]
so that $[P+Q] = (\varphi_x(x),\varphi_y(x))$ for every point $P = (x,v(x))$ of $D$. The pair $(\varphi_x,\varphi_y)$ is thus a rational parametrization of the translate, but not yet its $(u,v)$-coordinates. Recovering them requires the characteristic polynomial of multiplication by $\varphi_x(X)$ in $\F_q[X]/u(X)$, followed by an interpolation; these two operations are where the term $\deg(D)^\omega$ in the complexity appears (Remarks~\ref{rem:charpol} and~\ref{rem:interpolation}).

A difficulty is that the translate of a simple divisor need not be simple: it may happen that $[P+Q] = [-P'-Q]$ for two distinct points $P,P'$ of the support of $D$, in which case the two translates share an $x$-coordinate and no interpolant exists. Such collisions can be detected since they correspond to double roots of the characteristic polynomial. The algorithm interpolates the $y$-coordinates only modulo the factor of $u(X)$ whose roots are the $x$-coordinates of the non-colliding points: $\varphi_x$ is injective on these roots, so the interpolation problem is well posed. The colliding pairs are split off into separate simple summands, so the output is again a sum of simple divisors. The resulting procedure, Algorithm~\ref{algo:splitdivisortranslation}, translates a split divisor $D$ by a rational point within $\Otilde\bigl(\deg(D)\log^2 q + \deg(D)^\omega\bigr)$ operations in $\F_q$; see Proposition~\ref{prop:splitdivisortranslation}.

\subsection{The elliptic Mignotte--Schnorr recursion}\label{sec:ov_recursion}

Next, we explain how we separate points according to their torsion classes. Fix a prime $\ell$ dividing $N$, set $\mu_\ell := \val_\ell(N)$, write $N_\ell := N/\ell^{\mu_\ell}$ for the part coprime to $\ell$, and let $m \geq 1$. The aim of one level of the recursion is to split a divisor whose support lies in $E(\F_q)[N_\ell\cdot\ell^{m}]$ according to the class of its points in $E(\F_q)[\ell^m]/E(\F_q)[\ell^{m-1}]$. Given a system of representatives $J$ of that quotient, the test to be performed on a point $R$ of the support is whether $[R-P]\in E(\F_q)[N_\ell\cdot\ell^{m-1}]$, for each $P\in J$. This is a torsion membership test, and the division polynomial $\psi_{N_\ell\cdot\ell^{m-1}}(X,Y)$ cuts out exactly the corresponding points. Combining the translation formulas of Section~\ref{sec:ov_translation} with one GCD per class gives Algorithm~\ref{algo:EllipticCurveFactorizationAtLevelM}.

Rather than expanding the division polynomial, we evaluate $\psi_r(X,Y)$ directly at a generic point modulo $u(X)$ using its recurrence relations, in $\Otilde(\log r \cdot \deg u)$ operations (Remark~\ref{rem:computing_psi}); this is the analog of binary exponentiation in the cyclotomic case. The generic points to which the test is applied are the translates provided by Section~\ref{sec:ov_translation}. Algorithm~\ref{algo:EllipticCurveFactorizationAtLevelM} performs $\Otilde(\ell^2\deg(D)\log q)$ operations in $\F_q$, and only $\Otilde(\ell\deg(D)\log q)$ when the quotient $E(\F_q)[\ell^m]/E(\F_q)[\ell^{m-1}]$ is cyclic; see Proposition~\ref{prop:ECFactoAtLevelM}.

\paragraph{Rank-two torsion.}
The factor $\ell^2$ has no cyclotomic counterpart. Since $\F_q^\times$ is cyclic, its torsion quotients have exactly $\ell$ classes, and one level of Mignotte--Schnorr costs $\ell$ GCDs. The group $E(\F_q)$, by contrast, has rank up to $2$, so $E(\F_q)[\ell^m]/\allowbreak E(\F_q)[\ell^{m-1}]$ may be isomorphic to $(\Z/\ell\Z)^2$ and $\lvert J\rvert = \ell^2$. This is why $\Xi$ is squared in Theorem~\ref{thm:intro-rootfinding}, and why Remark~\ref{rem:cyclic_rootfinding} recovers a linear dependence when both groups are cyclic. The set $J$ itself is read off the datum: with $\nu_i := \val_\ell(\order(G_i))$,
\[
J = \bigl\{\,[\lambda_1\cdot \ell^{\nu_1-m}N_\ell\cdot G_1 + \lambda_2\cdot\ell^{\nu_2-m}N_\ell\cdot G_2] \;:\; \lambda_1,\lambda_2\in\{0,\ldots,\ell-1\}\,\bigr\},
\]
the term in $G_i$ being omitted when $m>\nu_i$.

Iterating over $m$, with a translation between consecutive levels, decomposes $D$ along the whole $\ell$-Sylow subgroup; this is Algorithm~\ref{algo:ellipticcurveprimaryfacto}, which performs $\Otilde\bigl(\mu_\ell(\ell^2\deg(D)\log q+\deg(D)\log^2 q+\deg(D)^\omega)\bigr)$ operations in $\F_q$, again with $\ell$ in place of $\ell^2$ when $E(\F_q)[\ell^\infty]$ is cyclic; see Proposition~\ref{prop:ellipticcurveprimaryfacto}.

Running this for every prime $\ell$ dividing $N$ separates any two distinct points of the support: if $P\neq P'$ lie in $E(\F_q)$, some prime power in the group order distinguishes them. Hence every simple divisor is eventually reduced to degree $1$, and a divisor of degree $1$ with coordinates $(u,v)$ gives back the root via $u(0)$. Doing this on $E$ and on $\widetilde E$ yields Algorithm~\ref{algo:ellipticcurverootfinding}.

\begin{example}[Running example: one level of the recursion]\label{ex:torsion83}
Continuing Example~\ref{ex:lift83}, take $\ell = 2$. Since $E(\F_q)\simeq\Z/2\Z\times\Z/2^2\Z\times\Z/3^2\Z$, the image of multiplication by $18$ is isomorphic to $\Z/2\Z$. Computing $\psi_{18}(X,Y)/(2Y)$, a polynomial in $X$ since $18$ is even, modulo $f_E(X)$ by the recurrences of Remark~\ref{rem:computing_psi} gives
\[
\psi_{18}(X,Y)/(2Y) \equiv 44X^4 + 21X^3 + 36X^2 + 32X + 2 \bmod f_E(X).
\]
As $\GCD(f_E,h) = 1$, the roots of $f_E$ that are $x$-coordinates of $18$-torsion points are the common roots of $f_E$ and of this quotient, whence $f_E = f_{E,0}\cdot f_{E,1}$ with
\[
f_{E,0}(X) = \GCD\bigl(f_E(X),\ \psi_{18}(X,Y)/(2Y) \bmod f_E(X)\bigr) = X^3 + 49X^2 + 45X + 47.
\]
Iterating over the levels and over $\ell\in\{2,3\}$, with a translation between consecutive levels, drives every factor down to degree $1$.
\end{example}

\subsection{Assembling the complexity bound}\label{sec:ov_complexity}

Table~\ref{tab:costs} sums up the complexity bounds established above. It remains to explain how the last line follows from the others, since this is the only point at which the quantity $\Xi$ appears.

\begin{table}[ht]
\centering
\renewcommand{\arraystretch}{1.3}
\begin{tabular}{@{}llll@{}}
\toprule
\textbf{Subroutine} & \textbf{Algorithm} & \textbf{Cost in operations in $\F_q$} & \textbf{Reference} \\
\midrule
$2$-adic splitting of $f$
  & Alg.~\ref{algo:TwoadicFactoWithEval}
  & $\Otilde\bigl(n\log^2 q\bigr)$
  & Prop.~\ref{prop:twoadicfactowitheval} \\
Lifting onto $E$ and $\widetilde E$
  & Alg.~\ref{algo:curveembedding}
  & $\Otilde\bigl(n\log^2 q\bigr)$
  & Prop.~\ref{prop:finitefieldembedding} \\
Translating a split divisor $D$
  & Alg.~\ref{algo:splitdivisortranslation}
  & $\Otilde\bigl(\deg(D)\log^2 q+\deg(D)^\omega\bigr)$
  & Prop.~\ref{prop:splitdivisortranslation} \\
One level of the recursion
  & Alg.~\ref{algo:EllipticCurveFactorizationAtLevelM}
  & $\Otilde\bigl(\ell^2\deg(D)\log q\bigr)$
  & Prop.~\ref{prop:ECFactoAtLevelM} \\
Decomposition along the $\ell$-Sylow
  & Alg.~\ref{algo:ellipticcurveprimaryfacto}
  & \begin{tabular}[t]{@{}l@{}}$\Otilde\bigl(\mu_\ell(\ell^2\deg(D)\log q$\\ $\qquad{}+\deg(D)\log^2 q+\deg(D)^\omega)\bigr)$\end{tabular}
  & Prop.~\ref{prop:ellipticcurveprimaryfacto} \\
\midrule
Whole online phase
  & Alg.~\ref{algo:ellipticcurverootfinding}
  & $\Otilde\bigl(\Xi^2 n\log^2 q+n^\omega\log q+n\log^3 q\bigr)$
  & Thm.~\ref{thm:rootfinding} \\
\bottomrule
\end{tabular}
\caption{Costs of the subroutines of the online phase, with $n = \deg(f)$.}
\label{tab:costs}
\end{table}

Recall that the lifting splits $f = u_0u_1$, and that the recursion is then run on $E$, over the primes $\ell\mid N$, for the roots of $u_0$, and on $\widetilde E$, over the primes $\ell\mid\widetilde N$, for those of $u_1$. We carry out the analysis for $E$. It calls Algorithm~\ref{algo:ellipticcurveprimaryfacto} once per prime $\ell\mid N$, on divisors whose degrees sum to $\deg(u_0)$; since $\omega\geq 2$, the costs add up at worst as if a single divisor of degree $\deg(u_0)$ were processed. The total is therefore $\Otilde\bigl(\sum_{\ell\mid N} \mu_\ell\,(\ell^2\deg(u_0)\log q+\deg(u_0)^\omega+\deg(u_0)\log^2 q)\bigr)$, and the whole dependence on the field is in the sums $\sum_\ell \mu_\ell \ell^2$ and $\sum_\ell \mu_\ell$. By definition of $\Xi$, every $\ell$ dividing $N$ satisfies $\ell\leq\Xi$, and since $\prod_\ell \ell^{\mu_\ell} = N \leq q+1+2\sqrt q$, we have $\sum_{\ell\mid N}\mu_\ell \leq \log_2 N = O(\log q)$. Hence $\sum_\ell \mu_\ell\ell^2 \leq \Xi^2\sum_\ell\mu_\ell = O(\Xi^2\log q)$. A similar analysis on $\widetilde E$ gives the same bound with $\deg(u_1)$ in place of $\deg(u_0)$. Adding the two runs and the cost of the lifting yields $\Otilde\bigl(\Xi^2 n\log^2 q + n^\omega\log q + n\log^3 q\bigr)$, which is Theorem~\ref{thm:intro-rootfinding}. The same computation with $\ell$ in place of $\ell^2$ when both groups are cyclic gives the improved bound of Remark~\ref{rem:cyclic_rootfinding}.

\subsection{The offline phase}\label{sec:ov_offline}

The existence and construction of good preprocessing data were discussed in Section~\ref{sec:intro_contrib}, and are treated in detail in Section~\ref{sec:precomp}; we only explain here why a datum is easy to verify. Besides elementary checks on $\beta$, $E$, $\widetilde E$ and the $G_i$, and trial division up to $\Xi$ for the claimed factorizations of $N$ and $\widetilde N$, it suffices to check that $G_1$ and $G_2$ are independent and that $\order(G_1)\order(G_2) = N$. Then $N$ divides $\lvert E(\F_q)\rvert$, and for $q\geq 37$ the Hasse--Weil interval contains no other multiple of $N$: the group order is certified without any point counting. With the same checks on $\widetilde E$, the verification is deterministic, unconditional, and polynomial in $\log q$ and $\Xi$ (Proposition~\ref{prop:verification}).

\subsection{Organization of the paper}\label{sec:ov_organization}
Section~\ref{sec:prelim} fixes notation and recalls the necessary background on elliptic curves and division polynomials. Sections~\ref{sec:lifting} and~\ref{sec:main_algo} are devoted to the online phase, and prove the statements of Sections~\ref{sec:ov_lifting}--\ref{sec:ov_complexity}: Section~\ref{sec:lifting} shows how to embed the roots into rational points, and Section~\ref{sec:main_algo} presents the decomposition of split divisors and our main algorithm. Section~\ref{sec:precomp} turns to the offline phase, and establishes in Section~\ref{sec:verification} that a datum can be verified efficiently. Section~\ref{sec:expe} provides experimental evidence for Conjecture~\ref{conj:main_conj} and describes our proof-of-concept implementation. Section~\ref{sec:conclusion} concludes with perspectives and open questions. Appendix~\ref{sec:app_MS} restates the Mignotte--Schnorr algorithm in the formalism of this paper.

\paragraph{Acknowledgements.} The author thanks Bill Allombert, Alin Bostan, Pierrick Gaudry, Joris van der Hoeven, François Morain, Christophe Petit, Benjamin Smith for helpful discussions. This work received funding from the France 2030 program managed by the French National Research Agency under grant agreement No. ANR-22-PETQ-0008 PQ-TLS.

\paragraph{AI methodology.} We used Claude Opus 5.5 (Anthropic) to assist with the presentation and the wording, the detection of typos and minor errors, and a few scientific contributions. The tool materially affected Sections~\ref{sec:distrib_smooth} and~\ref{sec:expe_conj}, as well as the complexity analyses of Sections~\ref{sec:lifting} and~\ref{sec:main_algo}. In more detail, the scientific project underlying this work was carried out without any AI tool: a complete earlier version of this paper, together with the accompanying Magma code, was written without any such tool, and Claude Opus 5.5 was only used afterwards. Compared to the first AI-free version of this work, a few scientific contributions are to be credited to Claude Opus 5.5: the connection between Conjecture~\ref{conj:main_conj} and the question of Erd\H os and Graham on sums of two smooth integers, and the heuristic estimate based on the Dickman function that supports this conjecture (Sections~\ref{sec:distrib_smooth} and~\ref{sec:expe_conj}). Some improvements of the complexity bound were also suggested by AI: the AI-free version of the complexity was $\widetilde O\bigl(\Xi(\mathcal D)^2\,(n^\omega+n\log^2 q)\bigr)$, and the AI suggested the optimizations required to obtain the improved bound. All passages rephrased with the help of AI have been carefully proofread by the author, and modified where needed, to ensure that they convey the intended meaning. The author verified the correctness and originality of all content, including references.

\section{Preliminaries and notation}\label{sec:prelim}

\paragraph{General notation.} Throughout the paper, $p$ denotes a prime greater than $3$, and $q = p^e$ is a power of $p$. When a finite field $\F_q$ is given as input to an algorithm, we implicitly assume that an $\F_p$-basis of $\F_q$ is provided, together with its multiplication table.
The assumption $p > 3$ allows us to use short Weierstrass equations as convenient models for elliptic curves.

For real numbers $x \leq y$, we denote by $\intrange{x}{y} = [x,y]\cap\Z$ the set of integers between $x$ and $y$. For an integer $n$, we write $\odd(n) := n / 2^{\val_2(n)}$.
We write $\omega \in [2,3]$ for a feasible exponent of matrix multiplication; in particular, $\omega < 2.371339$ using~\cite{alman2025more}.

We occasionally assume the Generalized Riemann Hypothesis (GRH); whenever this assumption is used, it will be stated explicitly.

We use the standard notations $O(\cdot)$ and $\Otilde(\cdot)$. Let $\varphi_1,\varphi_2 : \mathcal X \to \R_{\ge 0}$ be functions on some domain $\mathcal X$. We write $\varphi_1 = O(\varphi_2)$ (or $\varphi_1 \le O(\varphi_2)$) if there exists a constant $C > 0$ such that for all $x \in \mathcal X$, $\varphi_1(x) \le C\cdot \varphi_2(x)$. We write $\varphi_1 = \Otilde(\varphi_2)$ if there exist constants $C, \kappa > 0$ such that for all $x \in \mathcal X$, $\varphi_1(x) \le C\, \varphi_2(x)\, \log(\exp(1) + \varphi_2(x))^\kappa$.

Finally, we recall that the \emph{exponent} of an abelian group is the maximum order of its elements.

\paragraph{Elliptic curves and their quadratic twists.}
An elliptic curve is a smooth projective curve of genus $1$, with a fixed rational point. For general background, we refer to the standard references~\cite{silverman2009arithmetic, cohen2005handbook}.

When $p > 3$, any elliptic curve $E$ over $\F_q$ admits a short Weierstrass model: it is $\F_q$-isomorphic to the projective closure of an affine curve given by an equation $Y^2 = X^3 + aX + b$, for some $a,b \in \F_q$ satisfying $4a^3 + 27b^2 \ne 0$. The $j$-invariant of such a curve is defined by $j(E) := 6912\, a^3/(4a^3 + 27b^2)$. Two elliptic curves defined over $\F_q$ are $\overline{\F_q}$-isomorphic if and only if they have the same $j$-invariant.

Let $E$ be given by a short Weierstrass equation $Y^2 = X^3 + aX + b$, and let $\beta \in \F_q^\times$ be a nonsquare. The quadratic twist of $E$ by $\beta$ is the elliptic curve over $\F_q$ defined by the projective closure of $\beta Y^2 = X^3 + aX + b$.

Over finite fields, all quadratic twists of $E$ are $\F_q$-isomorphic. In particular, a representative can be written in short Weierstrass form as $Y^2 = X^3 + a\beta^{-2} X + b\beta^{-3}$, and we denote this curve by $\widetilde E$.

\paragraph{Rational points on elliptic curves.}
An elliptic curve given as the projective closure of a short Weierstrass equation has a unique point at infinity, denoted by $\infty$. The set of rational points $E(\F_q)$ consists of the affine points with coordinates in $\F_q$, together with $\infty$.

The set $E(\F_q)$ carries a group structure, where $\infty$ is the identity element. This group law can be described geometrically: if a projective line intersects $E$ at three rational points (counted with multiplicity), then their sum equals $\infty$.

The integer $U := q+1-\lvert E(\F_q)\rvert$ is the (Frobenius) trace of $E$. By Hasse's theorem~\cite[Thm.~V.1.1]{silverman2009arithmetic}, $\lvert U\rvert\leq 2\sqrt q$, i.e. $\lvert E(\F_q)\rvert$ lies in the Hasse--Weil interval $\intrange{q+1-2\sqrt q}{q+1+2\sqrt q}$. The quadratic twist $\widetilde E$ has trace $-U$, that is, $\lvert E(\F_q)\rvert+\lvert\widetilde E(\F_q)\rvert = 2(q+1)$: in the model $\beta Y^2 = X^3+aX+b$, each $x\in\F_q$ is the abscissa of exactly two affine points of $E$ and $\widetilde E$ together.

It is well-known that $E(\F_q)$ is isomorphic to $\Z/m_1\Z \times \Z/m_2\Z$ for integers $m_1, m_2$ such that $m_2 \mid m_1$~\cite[Sec.~6]{miller2004weil}. 
Given points $P_1, \ldots, P_r \in E(\F_q)$ and integers $n_1, \ldots, n_r \in \Z$, we denote by $\bigl[\sum_{i=1}^{r} n_i P_i\bigr]$ the point of $E(\F_q)$ obtained by applying the group law. The brackets are used to distinguish this point from the corresponding formal sum viewed as a divisor.

For a positive integer $n$, we denote by $E(\F_q)[n]$ the $n$-torsion subgroup, that is, the set of points $P \in E(\F_q)$ such that $[nP] = \infty$. In particular, if $\ell$ is prime, then $E(\F_q)[\ell]$ is either trivial (if $\ell \nmid m_1$), isomorphic to $\Z/\ell\Z$ (if $\ell \mid m_1$ but $\ell \nmid m_2$), or isomorphic to $\Z/\ell\Z \times \Z/\ell\Z$ (if $\ell \mid m_2$).

More generally, for $i \geq 1$, the quotient $E(\F_q)[\ell^i] / E(\F_q)[\ell^{i-1}]$ is either trivial, isomorphic to $\Z/\ell\Z$, or isomorphic to $\Z/\ell\Z \times \Z/\ell\Z$, depending on the $\ell$-adic valuations of $m_1$ and $m_2$.

We denote by $E(\F_q)[\ell^\infty]$ the $\ell$-Sylow subgroup of $E(\F_q)$, that is, the set of points whose order is a power of $\ell$. It is isomorphic to $\Z/\ell^{\val_\ell(m_1)}\Z \times \Z/\ell^{\val_\ell(m_2)}\Z$.

Finally, for an elliptic curve $E$ over $\F_q$ with quadratic twist $\widetilde E$, we denote by $\Xi(E)$ the largest prime factor of the product $\lvert E(\F_q)\rvert\cdot\lvert \widetilde E(\F_q)\rvert$. This notation is consistent with that of Definition~\ref{def:datum}: if $E$ is the curve of an elliptic preprocessing datum $\mathcal D$, then $\Xi(\mathcal D) = \Xi(E)$.

\paragraph{Division polynomials.}
Let $E$ be an elliptic curve over $\F_q$ given by a short Weierstrass equation $Y^2 = X^3 + aX + b$, and let $r \geq 1$ be an integer. The $r$-th division polynomial $\psi_r(X,Y) \in \F_q[X,Y]$ is a polynomial whose zero locus (in $\overline{\F_q}$) corresponds to the nonzero $r$-torsion points of $E$.
These polynomials satisfy recurrence relations that are central to fast computations (see Remark~\ref{rem:computing_psi}). We refer to~\cite[Sec.~4.4.5.a]{cohen2005handbook} for further details.

\section{Encoding the roots as rational points on an elliptic curve}\label{sec:lifting}

A split separable polynomial $f(X)$ defines an ideal $\langle f(X)\rangle \subset \F_q[X]$. An elliptic curve $E$ given by a Weierstrass equation $Y^2 = h(X)$ gives rise to a degree-$2$ extension of rings
\[
\iota : \F_q[X] \hookrightarrow \F_q[E] := \F_q[X,Y]/(Y^2 - h(X)),
\]
where $\F_q[E]$ denotes the ring of functions regular on $E \setminus \{\infty\}$.

The goal of this section is to describe a fast deterministic algorithm to construct, when possible, generators of an ideal $I \subset \F_q[E]$ such that $\iota$ induces an isomorphism
\[
\F_q[X]/(f(X)) \simeq \F_q[E]/I.
\]
Equivalently, we require that $\iota^{-1}(I) = \langle f(X)\rangle$ and that
\[
\dim_{\F_q}(\F_q[X]/(f(X))) = \dim_{\F_q}(\F_q[E]/I).
\]

In other words, we seek to lift the roots of $f$ to rational points on the elliptic curve $E$.
This is possible only if $h(\alpha)$ is a square in $\F_q$ for every root $\alpha$ of $f$: composing the isomorphism above with the evaluation at $\alpha$ yields a ring morphism $\F_q[E] \to \F_q$ sending $X$ to $\alpha$, and the image $y$ of $Y$ then satisfies $y^2 = h(\alpha)$, so that $(\alpha, y) \in E(\F_q)$.
Conversely, when this condition holds, the following lemma shows that such an ideal $I$ exists, namely the ideal generated by $f(X)$ and $Y-v(X)$, where $v(X) \in \F_q[X]$ has degree less than $\deg(f)$ and satisfies $v(X)^2 \equiv h(X) \bmod f(X)$.

\begin{lemma}\label{lem:separateEtwist}
  \begin{enumerate}
  \item If $h(\alpha)$ is a square in $\F_q$ for every root $\alpha$ of $f(X)$, then there exists a (non-unique) polynomial $v(X) \in \F_q[X]$ of degree less than $\deg(f)$ such that $v(X)^2 \equiv h(X) \bmod f(X)$.
  \item Let $\beta$ be a nonsquare in $\F_q^\times$. If $h(\alpha)$ is a nonsquare in $\F_q$ for every root $\alpha$ of $f(X)$, then there exists a (non-unique) polynomial $v(X) \in \F_q[X]$ of degree less than $\deg(f)$ such that $\beta\, v(X)^2 \equiv h(X) \bmod f(X)$.
  \end{enumerate}
\end{lemma}

\begin{proof}
We prove the first statement. For each root $\alpha$ of $f(X)$, choose a square root $\widehat{\alpha} \in \F_q$ of $h(\alpha)$. By interpolation, there exists a polynomial $v(X)$ of degree less than $\deg(f)$ such that $v(\alpha) = \widehat{\alpha}$ for all roots $\alpha$ of $f(X)$. This polynomial satisfies $v(X)^2 \equiv h(X) \bmod f(X)$.

The second statement is proved similarly, by choosing for each root $\alpha$ a square root $\widehat{\alpha} \in \F_q$ of $\beta^{-1} h(\alpha)$.
\end{proof}

\subsection{\texorpdfstring{$2$}{2}-adic factorization via evaluation}

Assume now that $f(X)$ is coprime with $h(X)$, so that $h(\alpha) \ne 0$ for every root $\alpha$ of $f(X)$. Fix a nonsquare $\beta \in \F_q^\times$. We compute a factorization $f(X) = \prod_{j \in S} f^{(j)}(X)$, where $S \subset \intrange{0}{2^{\val_2(q-1)}-1}$ is finite, and such that for every root $\alpha$ of $f^{(j)}(X)$, the element $h(\alpha) \beta^{-j}$ has odd multiplicative order. As will follow from Lemma~\ref{lem:TonelliShanks}, $h(\alpha)$ is then a square if and only if $j$ is even. Each factor $f^{(j)}(X)$ with $j$ even (resp.~odd) therefore satisfies the conditions of the first item (resp.~second item) of Lemma~\ref{lem:separateEtwist}, which allows us to lift its roots to rational points on $E$ (resp.~on its quadratic twist $\widetilde E$).

We now describe the core subroutine underlying this factorization, which is the
Mignotte--Schnorr recursion for
$\ell = 2$, applied to the values $h(\alpha)$ rather than to the roots $\alpha$
themselves.

Suppose first that we only wish to split $f(X)$ into two factors $f_0(X)$ and
$f_1(X)$, where $f_0$ (resp.~$f_1$) groups the roots $\alpha$ such that
$h(\alpha)$ is a square (resp.~a nonsquare). Since $h(\alpha)^{(q-1)/2} = \pm 1$
for every root $\alpha$ of $f(X)$, with the sign recording whether $h(\alpha)$ is
a square, it suffices to compute
\[
g(X) := h(X)^{(q-1)/2} \bmod f(X)
\]
by binary exponentiation, and then to set $f_i(X) := \GCD(g(X)-(-1)^i, f(X))$ for
$i\in\{0,1\}$.

To refine this into the full $2$-adic decomposition, we recurse on each $f_i(X)$,
halving the exponent at each level: for the recursive call with parameter $m$, the relevant
exponent is
$(q-1)/2^{\val_2(q-1)-m+1}$, and the value $h(\alpha)$ is changed to
$h(\alpha)\beta^{-i}$ so that the recursive call inputs an element whose order has
$2$-adic valuation at most $m-1$. The recursion has depth at most
$\val_2(q-1) \le \log_2 q$, and the degrees of the $f_i$ sum to $\deg(f)$ at each
level, which is what yields the bound of
Proposition~\ref{prop:twoadicfactowitheval}.
The overall procedure is summarized in Algorithm~\ref{algo:TwoadicFactoWithEval}.

\begin{algorithm}
  \caption{\TwoadicFactoWithEval{}: $2$-adic factorization via evaluation. \label{algo:TwoadicFactoWithEval}
  }
    \INPUT{
  \begin{itemize}[algo]
    \item A prime power $q$;
    \item A nonnegative integer $m\leq \val_2(q-1)$;
    \item A univariate polynomial $h(X)\in\F_q[X]$;
    \item A monic, split and separable polynomial $f(X)\in\F_q[X]$ coprime with $h(X)$ such that every root $x$ of $f(X)$ satisfies $\val_2(\order(h(x)))\leq m$;
    \item An element $\beta\in\F_q^\times$ such that $\val_2(\order(\beta)) = m$.
  \end{itemize}
    }
    \OUTPUT{
  \begin{itemize}[algo]
      \item A finite subset $S\subset \intrange{0}{2^m-1}$;
      \item A finite set of pairs $\{(j, f^{(j)}(X))\}_{j\in S}$ such that 
      $f^{(j)}(X)\in\F_q[X]$ are monic nonconstant polynomials, $\prod_{j\in S} f^{(j)}(X) = f(X)$ and
      for $j\in S$, 
      $$f^{(j)}(X) =
      \GCD(f(X),\prod_{\substack{\zeta\in\F_q, h(\zeta)\ne 0\text{ and}\\\order(h(\zeta)\cdot
      \beta^{-j})\text{ is odd}}}(X-\zeta)).$$
      \end{itemize}
    }
    \lIf{$f(X) = 1$\label{line:twoadicfactowitheval:casedegf0}}{
      Return $\{\}, \{\}$
    }
    \lIf{$m=0$\label{line:twoadicfactowitheval:casem0}}{
      Return $\{0\}, \{(0, f(X))\}$
    }
    $S\gets\{\}$\;
    $g(X)\gets h(X)^{(q-1)/2^{\val_2(q-1)-m+1}}\bmod f(X)$\label{line:twoadicfactowitheval:g}\;

    \For{$i\in\{0, 1\}$}{
      $f_i(X) \gets \GCD(g(X)-(-1)^i, f(X))$\label{line:twoadicfactowitheval:gcd_fi}\;
      $S_i, (j, f_i^{(j)})_{j\in S_i}\gets \TwoadicFactoWithEval(q, m-1, \beta^{-i}(h(X) \bmod f_i(X)), f_i(X), \beta^2)$\label{line:twoadicfactowitheval:recur4}\;
      $S\gets S \cup \{2\cdot j+i : j\in S_i\}$\;
      \lFor{$j\in S_i$}{
        $f^{(2\cdot j+i)}(X) \gets f_i^{(j)}(X)$
      }
    }
    Return $S, \{(j, f^{(j)}(X))\}_{j\in S}$\label{line:twoadicfactowitheval:return4}.
\end{algorithm}

\begin{proposition}\label{prop:twoadicfactowitheval}
  Algorithm~\ref{algo:TwoadicFactoWithEval} (\TwoadicFactoWithEval{}) terminates and is correct. It performs at most $\Otilde(m(\deg(f)\log q+\deg(h)))$ arithmetic operations in $\F_q$.
\end{proposition}

\begin{proof}
Termination follows immediately, since each recursive call decreases $m$ by one.

\smallskip

We next prove correctness. The return statements at Lines~\ref{line:twoadicfactowitheval:casedegf0} and \ref{line:twoadicfactowitheval:casem0} are clearly correct.

It remains to consider the return statement at Line~\ref{line:twoadicfactowitheval:return4}. We first show that the polynomials $f_0(X)$ and $f_1(X)$ computed at Line~\ref{line:twoadicfactowitheval:gcd_fi} satisfy
\begin{align}
  f_0(X) &= \prod_{\substack{\zeta \in \F_q \\ f(\zeta)=0 \\ \val_2(\order(h(\zeta))) <m}} (X-\zeta), \label{eq:f0}\\
  f_1(X) &= \prod_{\substack{\zeta \in \F_q \\ f(\zeta)=0 \\ \val_2(\order(h(\zeta))) =m}} (X-\zeta). \label{eq:f1}
\end{align}

Indeed, for every root $\zeta$ of $f(X)$, the equality
\[
h(\zeta)^{(q-1)/2^{\val_2(q-1)-m+1}}=1
\]
holds if and only if $\val_2(\order(h(\zeta)))<m$, whereas the equality
\[
h(\zeta)^{(q-1)/2^{\val_2(q-1)-m+1}}=-1
\]
holds if and only if $\val_2(\order(h(\zeta)))=m$.

We check that the inputs to the recursive calls at Line~\ref{line:twoadicfactowitheval:recur4} are valid.  The parameter passed in place of $m$ is $m-1$. The element passed in place of $\beta$ is $\beta^2$, and $\val_2(\order(\beta^2)) = \val_2(\order(\beta)) - 1 = m-1$ as required, because $\order(\beta)$ is even for $m\geq 1$. Finally, the polynomial passed in place of $h$ is $\beta^{-i}(h(X)\bmod f_i(X))$; by~\eqref{eq:f0} and~\eqref{eq:f1}, every root $\zeta$ of $f_i(X)$ satisfies $\val_2(\order(h(\zeta)))<m$ when $i=0$ and $\val_2(\order(h(\zeta)))=m$ when $i=1$, so that in both cases $\val_2(\order(\beta^{-i}h(\zeta)))\leq m-1$. Coprimality of $f_i(X)$ with this polynomial follows from that of $f(X)$ with $h(X)$.

By induction, each recursive call returns the required factorization. Combining the two outputs and replacing each index $j$ by $2j+i$ yields the required factorization of $f$. This is the same argument as in the classical Mignotte--Schnorr algorithm for $\ell=2$; see~\cite[Proof of Thm.~1]{mignotte1988calcul} or~\cite[Proof of Lem.~3]{grenet2015randomized}.

\smallskip

We now analyze the complexity. If $m=0$ or $f(X)=1$ then no arithmetic operation is performed. Otherwise, it computes $g(X)$ by binary exponentiation (Line~\ref{line:twoadicfactowitheval:g}), two GCDs (Line~\ref{line:twoadicfactowitheval:gcd_fi}) and the polynomials $\beta^{-i}(h(X)\bmod f_i(X))$. This costs $\Otilde(\deg(h)+\deg(f)\log q)$, that is, at most
\[
K\,(\deg(h)+\deg(f)\log q)\,\log\bigl(\exp(1)+\deg(h)+\deg(f)\log q\bigr)^\kappa
\]
operations in $\F_q$, for some constants $K,\kappa>0$ independent of the inputs.

We sum these costs over the tree of recursive calls. Let $d_h=\deg(h)$ and $d_f=\deg(f)$, and set $\Lambda := \log(\exp(1)+d_h+d_f\log q)$. Outside its recursive calls, the top-level call costs at most $K(d_h+d_f\log q)\Lambda^\kappa$. A call at depth $k\geq 1$ receives a factor $\widetilde f(X)$ of $f(X)$, together with a polynomial $\widetilde h(X)$ in place of $h(X)$. Since $\widetilde h(X)$ is obtained by reducing the polynomial of the parent call modulo $\widetilde f(X)$, we have $\deg(\widetilde h)<\deg(\widetilde f)$ and, by induction on the depth, $\deg(\widetilde h)\leq d_h$. Hence $\deg(\widetilde h)+\deg(\widetilde f)\log q$ is at most $d_h+d_f\log q$, which bounds the logarithmic factor by $\Lambda^\kappa$, and at most $2\deg(\widetilde f)\log q$ since $\log q\geq 1$. The cost of this call is therefore at most $2K\deg(\widetilde f)\log q\,\Lambda^\kappa$. The degrees of the factors $\widetilde f(X)$ received at a given depth sum to at most $d_f$ and only the depths $1,\ldots,m-1$ contribute. For $m\geq 1$, the total cost is therefore at most
\[
K(d_h+d_f\log q)\Lambda^\kappa + 2K(m-1)\,d_f\log q\,\Lambda^\kappa \;\leq\; 2K\,m\,(d_f\log q+d_h)\,\Lambda^\kappa.
\]
Finally, $\Lambda\leq\log\bigl(\exp(1)+m(d_f\log q+d_h)\bigr)$, which yields the bound $\Otilde(m(\deg(f)\log q+\deg(h)))$.

\end{proof}

\subsection{Lifting split polynomials to split divisors}

We begin by describing how to encode split divisors.
When a split divisor $D$ does not involve the point at infinity and the $x$-coordinates of the points in its support are distinct, it admits a rational parametrization, and it can be encoded using two univariate polynomials. This representation is closely related to the \emph{Mumford coordinates} used to encode points on Jacobians of hyperelliptic curves~\cite[Thm.~4.15]{cohen2005handbook}.

\begin{definition}\label{def:splitdivisor}
A \emph{split separable divisor} on $E$ is a finite formal sum $D = \sum_{i=1}^{d} P_i$, where the points $P_i \in E(\F_q)$ are pairwise distinct. If, in addition, $D$ does not contain the point at infinity and the $x$-coordinates of the points $P_i$ are pairwise distinct, then we say that it is \emph{simple}.

If $E$ is given by a short Weierstrass equation $Y^2 = X^3 + aX + b$, a simple divisor can be encoded by a pair $(u,v)$ of polynomials, where:
\begin{itemize}[noitemsep,nosep]
\item $u(X) \in \F_q[X]$ is the monic split polynomial whose roots are the $x$-coordinates of the points $P_i$;
\item $v(X) \in \F_q[X]$ is the unique polynomial of degree less than $\deg(u)$ such that $v(x_i) = y_i$ for all points $P_i = (x_i,y_i)$ in the support of $D$; in particular, $v(X)^2 \equiv X^3 + aX + b \bmod u(X)$.
\end{itemize}
\end{definition}

For simplicity, we abbreviate \emph{split separable divisor} as \emph{split divisor}.
Any split divisor $D$ can be written as $D = \sum_i D_i$ or $D = \infty + \sum_i D_i$, where the $D_i$ are simple divisors with disjoint supports; we encode $D$ by the $(u,v)$-coordinates of the $D_i$, together with a flag indicating whether $\infty$ lies in its support. This representation allows us to manipulate collections of points on $E$ using polynomial arithmetic.

\begin{remark}
The terminology \emph{split separable} reflects the fact that such divisors correspond to split separable ideals $I \subset \F_q[E] := \F_q[X,Y]/(Y^2 - X^3 - aX - b)$, i.e., ideals that decompose as intersections of maximal ideals with residue field $\F_q$. Equivalently, $\F_q[E]/I \simeq \F_q^m$ as an $\F_q$-algebra for some integer $m \ge 0$.
\end{remark}

We introduce the following elliptic-curve analogue of the root-finding problem:

\begin{problem}[Decomposition of split divisors on elliptic curves]\label{prob:ECdivdecomp}
Let $E$ be an elliptic curve over $\F_q$ given by a Weierstrass equation $Y^2 = h(X)$. Given two polynomials $u(X), v(X) \in \F_q[X]$ such that $u(X)$ is split and separable and $v(X)^2 \equiv h(X) \bmod u(X)$, compute the set of points $(x,y) \in \F_q^2$ satisfying $u(x)=0$ and $y = v(x)$.
\end{problem}

The goal of this section is to lift the classical root-finding problem to this elliptic-curve setting.

Split divisors play the role of split polynomials in the elliptic-curve setting. 
Algorithm~\ref{algo:curveembedding} lifts a split polynomial to a pair of split divisors $(D, \widetilde D)$, where $D$ lies on $E$ and $\widetilde D$ lies on its quadratic twist. This allows us to reduce the root-finding problem (Problem~\ref{prob:rootfinding}) to its elliptic-curve analogue (Problem~\ref{prob:ECdivdecomp}).

For a given $x \in \F_q$ such that $z := x^3 + ax + b$ is a nonzero square in $\F_q^\times$, the equation $Y^2 = z$ has exactly two solutions in $\F_q$. Equivalently, there are exactly two rational points on $E$ with $x$-coordinate equal to $x$. These points can be computed via modular square root extraction, for instance using the Tonelli--Shanks algorithm.

The next lemma provides a variant of the Tonelli--Shanks method, giving an explicit formula for a square root under the assumption that an exponent $j$ with $\order(z\beta^{-j})$ odd is known. This result will play a key role at Line~\ref{line:finitefieldembedding:algsqrt} of Algorithm~\ref{algo:curveembedding}.
The idea is to normalize $z$ using powers of a nonsquare $\beta$ so that its order becomes odd, allowing square roots to be extracted via exponentiation.

\begin{lemma}\label{lem:TonelliShanks}
Let $\beta, z \in \F_q^\times$ be such that $\beta$ is a nonsquare, and let $\gamma \in \Z$ be an inverse of $2$ modulo $\odd(q-1)$. Let $j \in \Z$ be such that $\order(z \beta^{-j})$ is odd, and define
\[
y := (z\beta^{-j})^\gamma \cdot \beta^{\lfloor j/2 \rfloor}.
\]
If $z$ is a square, then $j$ is even and $y$ is a square root of $z$ in $\F_q$. If $z$ is not a square, then $j$ is odd and $y$ is a square root of $z \beta^{-1}$.
\end{lemma}

\begin{proof}
The order of $w := z\beta^{-j}$ is odd and divides $q-1$, hence divides $\odd(q-1)$. Since $2\gamma \equiv 1 \bmod \odd(q-1)$, we get $w^{2\gamma} = w$, and therefore $y^2 = w\,\beta^{2\lfloor j/2\rfloor} = z\beta^{2\lfloor j/2\rfloor - j}$, which equals $z$ if $j$ is even and $z\beta^{-1}$ if $j$ is odd. Moreover, $w = (w^{\gamma})^2$ is a square; as $\beta$ is a nonsquare, $z = w\beta^{j}$ is a square if and only if $j$ is even.
\end{proof}

The simplest instance of Lemma~\ref{lem:TonelliShanks} occurs when $q \equiv 3 \bmod 4$. In this case, $j = 0$, $\beta = -1$, and $\gamma = (q+1)/4$. This recovers the classical map $x \mapsto x^{(q+1)/4}$, which sends squares to their square roots. This corresponds to the well-known exponentiation method for extracting square roots in finite fields when $q \equiv 3 \bmod 4$.

\begin{algorithm}
  \caption{\FiniteFieldEmbedding{}: lifting a split separable polynomial to split divisors on an elliptic curve and its quadratic twist.\label{algo:curveembedding}}
    \INPUT{
  \begin{itemize}[algo]
    \item A prime power $q$;
    \item A polynomial $h(X)\in\F_q[X]$;
    \item A monic, split and separable polynomial $f(X)\in\F_q[X]$ coprime with $h(X)$;
    \item A nonsquare $\beta$ in $\F_{q}^\times$.
  \end{itemize}
    }
    \OUTPUT{
      Four polynomials $u_0(X), u_1(X), v_0(X), v_1(X)\in\F_q[X]$ such that $u_0(X), u_1(X)$ are monic with $u_0(X)\cdot u_1(X) = f(X)$, and $\deg(v_0)<\deg(u_0)$, $\deg(v_1)<\deg(u_1)$, with $v_0^2(X)\equiv h(X)\bmod u_0(X)$ and $\beta\, v_1(X)^2\equiv h(X)\bmod u_1(X)$.
    }
    $S, \{(j, f^{(j)})\}_{j\in S}\gets \TwoadicFactoWithEval{}(q, \val_2(q-1), h(X), f(X), \beta)$\;
    $S_0\gets \{j : j\in S, j\text{ is even}\}$\;
    $S_1\gets \{j : j\in S, j\text{ is odd}\}$\;
    $u_0(X)\gets \prod_{j\in S_0} f^{(j)}(X)$\;
    $u_1(X)\gets \prod_{j\in S_1} f^{(j)}(X)$\;
    \tcc{$u_0$: roots mapping to $E$, $u_1$: roots mapping to the twist.}
    $\gamma\gets \text{ modular inverse of $2$ modulo }\odd(q-1)$\;
    \For{$j\in S$}{
      \tcc{Compute square roots via Lemma~\ref{lem:TonelliShanks}}
      $v^{(j)}(X)\gets (h(X)\beta^{-j})^\gamma \beta^{\lfloor j/2\rfloor}\bmod f^{(j)}(X)$\label{line:finitefieldembedding:algsqrt}\;
    }
    \tcc{Reconstruct global square roots via CRT}
    Using the CRT, compute $v_0(X)$ of degree less than $\deg(u_0(X))$ such that
    $v_0(X)\equiv v^{(j)}(X)\bmod f^{(j)}(X)$ for all $j\in S_0$\;
    Using the CRT, compute $v_1(X)$ of degree less than $\deg(u_1(X))$ such that
    $v_1(X)\equiv v^{(j)}(X)\bmod f^{(j)}(X)$ for all $j\in S_1$\;
    Return $u_0(X), u_1(X), v_0(X), v_1(X)$.
\end{algorithm}

We now analyze Algorithm~\ref{algo:curveembedding}, proving its correctness, termination, and complexity bounds.

\begin{proposition}\label{prop:finitefieldembedding}
Algorithm~\ref{algo:curveembedding} (\FiniteFieldEmbedding{}) terminates and is correct. It performs at most $\Otilde(\log(q)(\deg(f)\log q + \deg(h)))$ arithmetic operations in $\F_q$.
\end{proposition}

\begin{proof}
Termination follows directly from that of \TwoadicFactoWithEval{}, see Proposition~\ref{prop:twoadicfactowitheval}.

\smallskip

By the Chinese Remainder Theorem, it suffices to prove that the polynomials $v^{(j)}(X)$ computed at Line~\ref{line:finitefieldembedding:algsqrt} satisfy
\[
v^{(j)}(X)^2 \equiv h(X) \bmod f^{(j)}(X)
\]
when $j$ is even, and
\[
\beta\, v^{(j)}(X)^2 \equiv h(X) \bmod f^{(j)}(X)
\]
when $j$ is odd. This reduces to showing that for any root $\alpha$ of $f^{(j)}(X)$ (that is, any root $\alpha$ of $f(X)$ such that $(h(\alpha)\beta^{-j})^{\odd(q-1)} = 1$), we have
\[
v^{(j)}(\alpha)^2 = h(\alpha) \quad \text{(if $j$ is even)},
\]
and
\[
\beta\, v^{(j)}(\alpha)^2 = h(\alpha) \quad \text{(if $j$ is odd)}.
\]
This follows from Lemma~\ref{lem:TonelliShanks}, applied with $z = h(\alpha)$.

\smallskip

We now turn to the complexity analysis. Running \FiniteFieldEmbedding{} requires one call to \TwoadicFactoWithEval{}, one modular inversion, $\lvert S\rvert$ loop iterations, and one application of the Chinese Remainder Theorem. The cost of all these operations is straightforward to estimate. A slightly more delicate step is the computation of the values $\beta^{\lfloor j/2\rfloor - j\gamma}$, which can be performed using binary exponentiation at a total cost of $O(\deg(f)\log q)$, since $\lvert S\rvert \leq \deg(f)$.

The remainders $h(X)\bmod f^{(j)}(X)$, for $j\in S$, are computed beforehand: we first reduce $h(X)$ modulo $f(X)$, in $\Otilde(\deg(f)+\deg(h))$ operations, and then reduce the result simultaneously modulo all the $f^{(j)}(X)$ using a remainder tree, in $\Otilde(\deg(f))$ operations since $\sum_{j\in S}\deg(f^{(j)}) = \deg(f)$, see~\cite[Sec.~10.3]{von2003modern}. Given these remainders, each modular exponentiation at Line~\ref{line:finitefieldembedding:algsqrt} costs $\Otilde(\deg(f^{(j)})\log q)$ operations; summing over all $j$ yields a total cost of $\Otilde(\deg(f)\log q)$. Therefore, the complexity is dominated by the cost of the call to \TwoadicFactoWithEval{}, which is $\Otilde(\val_2(q-1)(\deg(f)\log q + \deg(h)))$ arithmetic operations by Proposition~\ref{prop:twoadicfactowitheval}. The inequality $\val_2(q-1)\leq \log q$ concludes the proof.
\end{proof}

\section{Factoring with elliptic curves}\label{sec:main_algo}

In this section, we study the Mignotte--Schnorr framework for solving the decomposition problem on elliptic curves (Problem~\ref{prob:ECdivdecomp}). This framework requires two operations: (i) separating points lying in the kernel of a multiplication map, and (ii) translating a divisor by a given point.

In the classical cyclotomic setting for factoring a polynomial $f(X)$, (i) is achieved by computing the GCD of $f(X)$ with $X^{r} - 1$ for suitable divisors $r$ of $q-1$, while (ii) is achieved by replacing $f(X)$ with $f(\gamma X)$ for some $\gamma \in \F_q^\times$.

In our elliptic-curve setting, when decomposing a simple divisor given by its $(u,v)$-coordinates, (i) is achieved by computing the GCD of $u(X)$ with the evaluations $\psi_r(X, v(X))$ of division polynomials, for suitable divisors $r$ of $\lvert E(\F_q)\rvert$. The goal of the next subsection is to describe an algorithm that realizes (ii).

\subsection{Translating split divisors}

In this subsection, we describe Algorithm~\ref{algo:splitdivisortranslation}, which provides a method to translate split divisors by a rational point $Q$ on the curve. The algorithm translates each simple component of the divisor by computing the image of its points under addition with $Q$, using algebraic operations on the $(u,v)$ representation. The main idea is to express the group law on $E$ in terms of polynomial operations, and to recover the translated divisor via characteristic polynomials.

\begin{algorithm}
  \caption{\SplitDivisorTranslation{}: Translation of a split divisor by a rational point.\label{algo:splitdivisortranslation}}
  \INPUT{
  \begin{itemize}[algo]
      \item An elliptic curve $E$ over $\F_q$ given by a short Weierstrass equation
        $Y^2 = X^3+aX + b$, with $a,b\in \F_{q}$;
      \item A split divisor encoded as a finite sum $D = \sum_{i\in I} D_i$ of simple divisors, where $D_i$ is encoded by its $(u,v)$-coordinates\;
      \item A rational point $Q\in E(\F_{q})$, written $Q=(x_Q, y_Q)$ when $Q\ne\infty$, such that neither $Q$ nor $-Q$ lie in the support of $D$;
      \item A nonsquare $\beta$ in $\F_{q}^\times$.
    \end{itemize}
  }
  \OUTPUT{
  \begin{itemize}[algo]
      \item A representation of the split divisor $\sum_{P\in D} [P+Q]$.
     \end{itemize}
   }
  \lIf{$Q=\infty$}{Return $D$\label{line:splitdivisortranslation:infty}}
  \For{$i\in I$}{
    Let $(u_i(X), v_i(X))$ be the $(u,v)$-coordinates of $D_i$\;
  $s_i(X)\gets (v_i(X)-y_Q)(X-x_Q)^{-1}\bmod u_i(X)$\;
  \tcc{$(X-x_Q)$ is invertible since $Q, -Q$ are not in the support.}
  $\varphi_i(X)\gets s_i(X)^2-X-x_Q\bmod u_i(X)$\;
  $\chi_i(X) \gets$ characteristic polynomial of the multiplication by $\varphi_i(X)$ in $\F_q[X]/u_i(X)$\label{line:translate_charpol}\;
  \tcc{See Remark~\ref{rem:charpol} on how to compute this characteristic polynomial.}
  $\widetilde u_{i, 2}(X)\gets \GCD(\chi_i(X), \frac{d}{dX}\chi_i(X))$\label{line:splitdivisortranslation:ui2}\;
  $\widetilde u_{i, 1}(X)\gets \chi_i(X)/\widetilde u_{i, 2}(X)^2$\label{line:splitdivisortranslation:ui1}\;
  \tcc{We have $\chi_i(X) = \widetilde u_{i,1}(X)\cdot \widetilde u_{i,2}(X)^2$.}
  $w_i(X)\gets \GCD(u_i(X), \widetilde u_{i,1}(\varphi_i(X))\bmod u_i(X))$\label{line:splitdivisortranslation:wi}\;
  \tcc{$w_i(X)$ is the product of the $(X - x_j)$ for those roots $x_j$ of $u_i(X)$ whose image under $\varphi_i$ is not shared by another root, and $\deg(w_i) = \deg(\widetilde u_{i,1})$.}
  Compute $\widetilde v_{i,1}(X)$ of degree less than $\deg(\widetilde u_{i,1})$ such that\label{line:splitdivisortranslation:interpolation} $$\widetilde v_{i,1}(\varphi_i(X))\equiv -y_Q+s_i(X)(x_Q-\varphi_i(X))\bmod w_i(X)\;$$
  \tcc{See Remark~\ref{rem:interpolation} for details.}
  Let $\widetilde D_{i,1}$ be the simple divisor with $(u,v)$-coordinates $\widetilde u_{i,1}(X), \widetilde v_{i,1}(X)$\;
  $\_, \_, \widetilde v_{i,2}(X), \_\gets\FiniteFieldEmbedding(q, X^3+aX+b, \widetilde u_{i,2}(X), \beta)$\label{line:splitdivisortranslation:ffembed}\;
  Let $\widetilde D_{i,2}$ and $\widetilde D_{i,3}$ be the simple divisors with respective coordinates $(\widetilde u_{i,2}(X), \widetilde v_{i,2}(X))$ and $(\widetilde u_{i,2}(X), -\widetilde v_{i,2}(X))$\;
}
  Return $\sum_{i\in I}(\widetilde D_{i,1} + \widetilde D_{i,2} + \widetilde D_{i,3})$.
\end{algorithm}

We begin with a few remarks on the subroutines used in this algorithm.

\begin{remark}\label{rem:interpolation}
To compute $\widetilde v_{i,1}(X)$ at Line~\ref{line:splitdivisortranslation:interpolation} in Algorithm~\ref{algo:splitdivisortranslation}, we need to solve the following interpolation problem. Given a split separable monic polynomial $u(X)$, and polynomials $g(X)$ and $\varphi(X)$ of degree less than $\deg(u)$, compute a polynomial $v(X)$ such that $v(\varphi(X)) \equiv g(X) \bmod u(X)$. This problem can be viewed as interpolating the values of $v(X)$ on the image of the roots of $u(X)$ under $\varphi(X)$. This is an implicit form of the Lagrange interpolation problem: the polynomial $v(X)$ must satisfy $v(\varphi(\alpha)) = g(\alpha)$ for every root $\alpha$ of $u(X)$. The difficulty lies in performing this interpolation without explicitly knowing the roots of $u(X)$.

For the problem to be well posed, we assume that $\varphi$ is injective on the roots of $u(X)$. Under this assumption, the conditions $v(\varphi(\alpha)) = g(\alpha)$, one for each root $\alpha$ of $u(X)$, form a square linear system whose unknowns are the $\deg(u)$ coefficients of $v(X)$, sought of degree less than $\deg(u)$. Its matrix is the Vandermonde matrix of the values $\varphi(\alpha)$, so $v(X)$ exists and is unique.
If the roots of $u(X)$ were known, such a polynomial $v(X)$ could be computed in $\Otilde(\deg(u))$ operations using a subproduct tree, see~\cite[Sec.~10.2]{von2003modern}. However, without access to the roots, the best method we are aware of is to solve a linear system whose unknowns are the coefficients of $v(X)$. Setting up this system requires the powers $\varphi(X)^k\bmod u(X)$ for $k<\deg(u)$, at a cost of $\Otilde(\deg(u)^2)$ operations, and solving it costs $O(\deg(u)^\omega)$. We discuss in Section~\ref{sec:improving_complexity} how this linear system might be avoided, by viewing the problem as an instance of inverse modular composition.
\end{remark}

\begin{remark}[Computing $\chi_i(X)$]\label{rem:charpol}
At Line~\ref{line:translate_charpol} of Algorithm~\ref{algo:splitdivisortranslation}, we need to compute the characteristic polynomial of the linear map
\[
\begin{array}{rcl}
\F_q[X]/u(X) &\longrightarrow& \F_q[X]/u(X)\\
f(X) &\longmapsto& f(X)\cdot \varphi(X) \bmod u(X),
\end{array}
\]
where $\deg(f) < \deg(u)$ and $u(X)$ is a split, separable, monic polynomial.
A matrix representation of this map in the basis $1, X, X^2, \ldots, X^{\deg(u)-1}$ can be computed in $O(\deg(u)^2)$ operations, by iteratively computing $X^i \varphi(X) \bmod u(X)$.
The characteristic polynomial of this matrix can then be computed deterministically in $O(\deg(u)^\omega)$ operations using the algorithm of Neiger and Pernet~\cite{neiger2021deterministic}, which removes the logarithmic factor in the complexity of Keller-Gehrig's algorithm~\cite{keller1985fast}. 
\end{remark}

\begin{proposition}\label{prop:splitdivisortranslation}
Algorithm~\ref{algo:splitdivisortranslation} (\SplitDivisorTranslation{}) terminates and is correct. It performs at most $\Otilde(\deg(D)\log^2 q + \deg(D)^\omega)$ arithmetic operations in $\F_q$.
\end{proposition}

\begin{proof}
Termination is immediate, since the algorithm involves no recursive calls and only finite loops. The termination of \FiniteFieldEmbedding{} at Line~\ref{line:splitdivisortranslation:ffembed} is established in Proposition~\ref{prop:finitefieldembedding}.

\medskip

We now prove correctness. The case $Q=\infty$ is settled at Line~\ref{line:splitdivisortranslation:infty}, since translation by the neutral element leaves every point of $D$ unchanged; we assume from now on that $Q=(x_Q,y_Q)$ is an affine point.
If $D$ is a sum of simple divisors, the algorithm processes each summand independently and returns the sum of the translated divisors. Therefore, it suffices to consider the case where $D = \sum_{j\in A} P_j$ is a simple divisor (so that $x(P_j) \ne x(P_{j'})$ for $j \ne j'$), encoded by its $(u,v)$-coordinates. In particular, $I$ contains a single index $i$.

A difficulty is that the translation of a simple divisor need not be simple. Define
\[
\widetilde A = \{j_1 \in A : \exists j_2 \in A \text{ such that } j_1\ne j_2\text{ and }[P_{j_1}+Q] = [-P_{j_2}-Q]\}.
\]
The set $\widetilde A$ encodes pairs of points whose images under translation by $Q$ have the same $x$-coordinate. Equivalently, $\sum_{j\in A} [P_j+Q]$ is a simple divisor if and only if $\widetilde A$ is empty.

We first show that
\[
\widetilde u_{i,1}(X) = \prod_{j\in A \setminus \widetilde A} (X - x([P_j+Q])), \qquad
\widetilde u_{i,2}(X) = \prod_{\xi\in \{x([P_j+Q])\,:\,j\in\widetilde A\}} (X - \xi).
\]
The second product runs over a set. Indeed, if $x([P_{j_1}+Q]) = x([P_{j_2}+Q])$, then $[P_{j_1}+Q] = \pm[P_{j_2}+Q]$, and the sign $+$ forces $j_1 = j_2$. Hence $\widetilde A$ is a disjoint union of pairs $\{j_1,j_2\}$ with $[P_{j_1}+Q] = [-P_{j_2}-Q]$, the two indices of a pair give the same abscissa, and $\widetilde u_{i,2}(X)^2 = \prod_{j\in \widetilde A} (X - x([P_j+Q]))$.
To prove these two identities, we show that
\[
\chi_i(X) = \prod_{j\in A} (X - x([P_j+Q])).
\]
For each $j\in A$, let $P_j = (x_j,y_j)$, where $x_j$ is a root of $u_i(X)$ and $y_j = v_i(x_j)$. Using the group law on elliptic curves (and the assumption that $P_j \ne Q$), we obtain $\varphi_i(x_j) = x([P_j+Q])$. Since
\[
(\varphi_i(X) - \varphi_i(x_j)) \cdot \left(\frac{u_i(X)}{X - x_j}\right) \equiv 0\bmod u_i(X),
\]
it follows that $\varphi_i(x_j)$ is an eigenvalue of multiplication by $\varphi_i(X)$ in $\F_q[X]/u_i(X)$. Thus, each $j \in A \setminus \widetilde A$ contributes an eigenvalue of multiplicity $1$, while each pair $\{j_1,j_2\}\subseteq\widetilde A$ as above contributes the eigenvalue $x([P_{j_1}+Q]) = x([P_{j_2}+Q])$ with multiplicity $2$ (corresponding to two eigenvectors). The computation of $\widetilde u_{i,2}(X)$ and $\widetilde u_{i,1}(X)$ at Lines~\ref{line:splitdivisortranslation:ui2} and~\ref{line:splitdivisortranslation:ui1} separates these contributions according to multiplicity.

Next, we show that for any $j \in A \setminus \widetilde A$, we have
\[
\widetilde v_{i,1}(x([P_j+Q])) = y([P_j+Q]).
\]
This follows from the relation
\[
\widetilde v_{i,1}(\varphi_i(X)) \equiv -y_Q + s_i(X)(x_Q - \varphi_i(X)) \bmod w_i(X).
\]
By construction, $w_i(X) = \prod_{j\in A\setminus\widetilde A}(X - x(P_j))$, so this congruence imposes exactly one condition at each $j\in A\setminus\widetilde A$, and $\varphi_i$ is injective on the roots of $w_i(X)$; by Remark~\ref{rem:interpolation}, $\widetilde v_{i,1}(X)$ is therefore well defined and unique.
Reducing this identity modulo $X - x(P_j)$, for $j\in A\setminus\widetilde A$, and using $\varphi_i(x(P_j)) = x([P_j+Q])$, together with the standard formula for the group law,
\[
y([P_j+Q]) = -y_Q + s_i(x_j)(x_Q - x([P_j+Q])),
\]
we conclude that $\widetilde v_{i,1}(X)$ provides the correct parametrization for 
$$\widetilde D_{i,1} = \sum_{j\in A\setminus\widetilde A}[P_j+Q].$$

Finally, we show that
\[
\widetilde D_{i,2} + \widetilde D_{i,3} = \sum_{j\in \widetilde A} [P_j+Q].
\]
By construction, the set $\{[P_j+Q]\}_{j\in \widetilde A}$ is stable under negation. We first prove that it contains no $2$-torsion point. Indeed, if $[P_j+Q]$ were a $2$-torsion point, then there would exist $j'\ne j$ such that
\[
[-P_j - Q] = [P_j+Q] = [-P_{j'} - Q],
\]
which would imply $P_j = P_{j'}$, contradicting $j \ne j'$.

Thus, \FiniteFieldEmbedding{} lifts the set $\{x([P_j+Q])\}_{j\in \widetilde A}$ to a simple divisor $\widetilde D_{i,2}$ containing exactly one point of each pair $\{R,-R\}$, for $R\in\{[P_j+Q]\}_{j\in\widetilde A}$. By construction, $\widetilde D_{i,3} = -\widetilde D_{i,2}$, and therefore
\[
\widetilde D_{i,2} + \widetilde D_{i,3} = \sum_{j\in \widetilde A} [P_j+Q],
\]
which concludes the proof of correctness.

\medskip

We now analyze the complexity. Within one iteration of the loop, computing $s_i(X)$ and $\varphi_i(X)$ costs $\Otilde(\deg(D_i))$, and so do the GCD and the division at Lines~\ref{line:splitdivisortranslation:ui2} and~\ref{line:splitdivisortranslation:ui1}. Four steps are more expensive:
\begin{itemize}[noitemsep,nosep]
\item computing the characteristic polynomial $\chi_i(X)$ at Line~\ref{line:translate_charpol} costs $O(\deg(D_i)^\omega)$ operations, see Remark~\ref{rem:charpol};
\item computing $w_i(X)$ at Line~\ref{line:splitdivisortranslation:wi} requires one modular composition $\widetilde u_{i,1}(\varphi_i(X))\bmod u_i(X)$ followed by one GCD. The modular composition costs $\Otilde(\deg(D_i)^{(\omega+1)/2})$ operations by the algorithm of Brent and Kung~\cite{brent1978fast} and the GCD costs $\Otilde(\deg(D_i))$, so this step is $O(\deg(D_i)^\omega)$, since $\omega\geq 2$;
\item computing $\widetilde v_{i,1}(X)$ at Line~\ref{line:splitdivisortranslation:interpolation} costs $O(\deg(D_i)^\omega)$ operations, see Remark~\ref{rem:interpolation};
\item calling \FiniteFieldEmbedding{} at Line~\ref{line:splitdivisortranslation:ffembed} costs $\Otilde(\deg(D_i)\log^2 q)$ operations, see Proposition~\ref{prop:finitefieldembedding}.
\end{itemize}

Summing over $i\in I$ and using $\sum_{i\in I}\deg(D_i) = \deg(D)$ together with the superadditivity of $n\mapsto n^\omega$, we obtain
\[
\sum_{i\in I} \Otilde\bigl(\deg(D_i)^\omega + \deg(D_i)\log^2 q\bigr) \le \Otilde\bigl(\deg(D)^\omega + \deg(D)\log^2 q\bigr)
\]
operations in $\F_q$, which is the claimed bound.
\end{proof}

\subsection{Deterministic decomposition of split divisors on elliptic curves}

At this point, we have all the subroutines needed to describe our deterministic factorization procedure. Before presenting the full algorithm, we first describe a partial factorization step that is applied for any prime $\ell$ dividing $\lvert E(\F_q)\rvert$, following the approach of Mignotte--Schnorr.

For a given prime $\ell$ and an integer $m$, Algorithm~\ref{algo:EllipticCurveFactorizationAtLevelM} decomposes a split divisor $D$ according to the projections of its points onto the quotient group $E(\F_q)[\ell^m] / E(\F_q)[\ell^{m-1}]$.

More precisely, let $P \in E(\F_q)$ be a point such that $\val_\ell(\order(P)) \le m$. Then $P$ can be projected onto $E(\F_q)[\ell^m]$ by multiplying it by the cofactor
\[
N_\ell = \frac{\lvert E(\F_q)\rvert}{\ell^{\val_\ell(\lvert E(\F_q)\rvert)}}.
\]
If $P$ and $Q$ are two such points for which $[N_\ell \cdot P]$ and $[N_\ell \cdot Q]$ lie in distinct classes modulo $E(\F_q)[\ell^{m-1}]$, then Algorithm~\ref{algo:EllipticCurveFactorizationAtLevelM} separates them, as they are assigned to distinct output divisors. Intuitively, this procedure performs one level of the separation of points according to their projection on the $\ell$-Sylow subgroup of $E(\F_q)$; the full separation is obtained by iterating it over the levels, see Algorithm~\ref{algo:ellipticcurveprimaryfacto} below. The algorithm combines translation on the elliptic curve with division polynomials to detect membership in torsion subgroups.

\begin{algorithm}
  \caption{\EllipticCurvePrimaryFactorizationAtLevelm{}:Partial factorization of a split divisor via projection onto $E(\F_q)[\ell^m]/E(\F_q)[\ell^{m-1}]$.\label{algo:EllipticCurveFactorizationAtLevelM}}
  \INPUT{
  \begin{itemize}[algo]
      \item An elliptic curve $E$ over $\F_{q}$ given by a short Weierstrass
        equation;
      \item A prime number $\ell$ dividing $\lvert E(\F_{q})\rvert$;
      \item A positive integer $m\leq \val_\ell(\exponent(E(\F_q)))$\;
      \item A set $J$ (of cardinality $\ell$ or $\ell^2$) of representatives for the classes in $E(\F_q)[\ell^m]/E(\F_q)[\ell^{m-1}]$\;
      \item A split divisor $D = \sum_{i\in I} D_i$, where each $D_i$ is a simple divisor, such that all points in its support lie in $E(\F_q)[N_\ell \cdot \ell^m]$, where $N_\ell = \lvert E(\F_q)\rvert/\ell^{\val_\ell(\lvert E(\F_q)\rvert)}$,
      and such that neither the support of $D$ nor its set of opposites intersects $J$.
    \end{itemize}
  }
  \OUTPUT{
      \begin{itemize}[algo]
      \item A family of split divisors $(D^{(P)})_{P\in J}$, where $D^{(P)}$ consists of the points $R \in D$ such that $[R-P] \in E(\F_q)[N_\ell \cdot \ell^{m-1}]$.
   \end{itemize}
  }
  \For{$i\in I$}{
    Let $(u_i(X), v_i(X))$ be the $(u,v)$-coordinates for $D_i$\;
    \For{$P\in J$}{
      \eIf{$P = \infty$\label{line:levelm:infty}}{
        $\varphi_{i, x}(X)\gets X$\;
        $\varphi_{i, y}(X)\gets v_i(X)$\;
      }{
      $s_i(X) \gets  (v_i(X)+y_P)(X-x_P)^{-1}\bmod u_i(X)$\;
        \tcc{$X - x_P$ is invertible modulo $u_i(X)$ since no point in the support of $D_i$ (nor its opposite) lies in $J$.}
      $\varphi_{i,x}(X)\gets s_i(X)^2 - X -x_P\bmod u_i(X)$\;
      $\varphi_{i,y}(X)\gets y_P +s_i(X)\cdot (x_P-\varphi_{i,x}(X))\bmod u_i(X)$\; 
      }
      $t_i(X) \gets \psi_{N_\ell\cdot \ell^{m-1}}(\varphi_{i,x}(X), \varphi_{i,y}(X))\bmod u_i(X)$\label{line:EllipticCurveFactorizationAtLevelM:divpol}\;
      \tcc{Computed directly in $\F_q[X]/u_i(X)$, see Remark~\ref{rem:computing_psi}.}
      $u_{i, P}(X)\gets \GCD(u_i(X), t_i(X))$\;
      $v_{i,P}(X)\gets v_i(X)\bmod u_{i,P}(X)$\; 
    }
    }
   Return $(\sum_{i\in I}D_{i,P})_{P\in J}$, where $D_{i,P}$ is the simple divisor with coordinates $(u_{i, P}(X), v_{i, P}(X))$.
\end{algorithm}

Algorithm~\ref{algo:EllipticCurveFactorizationAtLevelM} requires the efficient computation of modular reductions of division polynomials. The following remark explains how to perform this step.

\begin{remark}\label{rem:computing_psi}
For $r \in \Z_{>0}$ and a fixed elliptic curve $E$, the value $\psi_r(\varphi_x(X),\varphi_y(X)) \bmod u(X)$ of the division polynomial $\psi_r(X,Y)$ at a generic point $(\varphi_x(X),\varphi_y(X))$, as required at Line~\ref{line:EllipticCurveFactorizationAtLevelM:divpol} of Algorithm~\ref{algo:EllipticCurveFactorizationAtLevelM}, can be computed within $\Otilde(\log r \cdot \deg(u))$ operations, without expanding $\psi_r(X,Y)$.

Using the classical recurrence relations for division polynomials, we have
\[
\begin{array}{rcll}
\psi_{2r+1}(X,Y) &=& \psi_{r+2}(X,Y)\psi_r(X,Y)^3 - \psi_{r-1}(X,Y)\psi_{r+1}(X,Y)^3 & \text{for } r \ge 2, \\
2Y \cdot \psi_{2r}(X,Y) &=& \psi_r(X,Y)\bigl(\psi_{r+2}(X,Y)\psi_{r-1}(X,Y)^2 - \psi_{r-2}(X,Y)\psi_{r+1}(X,Y)^2\bigr) & \text{for } r \ge 3.
\end{array}
\]

These relations imply that any block of eight consecutive division polynomials starting at index $r$ can be computed from the block starting at index $\lceil r/2 \rceil - 2$ (base cases: $\psi_{-1} = -1$, $\psi_0 = 0$ and the explicit $\psi_1, \dots, \psi_4$). The second relation should not be used to divide by $2Y$, which may become a zero divisor modulo $u(X)$. Instead, we use the classical normalization $\psi_r(X,Y) = \Psi_r(X)$ for odd $r$ and $\psi_r(X,Y) = 2Y\,\Psi_r(X)$ for even $r$, where $\Psi_r(X)\in\F_q[X]$: substituting it into the relations above and replacing $Y^2$ by $X^3+aX+b$ yields recurrences on the polynomials $\Psi_r(X)$ that involve no division. These recurrences are polynomial identities, so they remain valid after substituting $\varphi_x(X)$ for $X$ and reducing modulo $u(X)$. We obtain a recursive algorithm computing $\Psi_r(\varphi_x(X)) \bmod u(X)$ with $O(\log r)$ operations in the quotient ring $\F_q[X]/(u(X))$, that is, $\Otilde(\log r \cdot \deg(u))$ arithmetic operations in $\F_q$. The requested value is $\Psi_r(\varphi_x(X))$ if $r$ is odd, and $2\varphi_y(X)\,\Psi_r(\varphi_x(X))$ if $r$ is even.

This approach is hinted at in~\cite[Sec.~3.1]{morain2007implementing}, see also~\cite[Proof of Lem.~4.1]{cheng2003straight} for a similar analysis. This is analogous to a fast doubling method for division polynomials, similar in spirit to binary exponentiation. We implemented it in our proof-of-concept software as the function \texttt{DivisionPolynomialAtPoint} in~\cite{implementation}.
\end{remark}

\begin{proposition}\label{prop:ECFactoAtLevelM}
  Algorithm~\ref{algo:EllipticCurveFactorizationAtLevelM} (\EllipticCurvePrimaryFactorizationAtLevelm{}) terminates and is correct. It performs at most $\Otilde(\ell^2\deg(D)\log q)$ operations in $\F_q$. This complexity bound can be improved to $\Otilde(\ell\deg(D)\log q)$ if $E(\F_q)[\ell^m]/E(\F_q)[\ell^{m-1}]$ is cyclic.
\end{proposition}

\begin{proof}
Termination is immediate, since the algorithm only involves loops over finite sets.

\medskip

We now prove correctness. We must show that the roots of $u_{i,P}(X)$ are precisely the $x$-coordinates of the points $R \in D_i$ such that $[R-P] \in E(\F_q)[N_\ell \cdot \ell^{m-1}]$.

First, an argument similar to the proof of Proposition~\ref{prop:splitdivisortranslation} shows that for any point $R \in D_i$, the coordinates of $[R-P]$ satisfy
\[
\varphi_{i,x}(x(R)) = x([R-P]), \quad \varphi_{i,y}(x(R)) = y([R-P]).
\]

Next, we observe that $[R-P] \in E(\F_q)[N_\ell \cdot \ell^{m-1}]$ if and only if
\[
\psi_{N_\ell \cdot \ell^{m-1}}(x([R-P]), y([R-P])) = 0,
\]
that is,
\[
\psi_{N_\ell \cdot \ell^{m-1}}(\varphi_{i,x}(x(R)), \varphi_{i,y}(x(R))) = 0.
\]

Finally, since $x(R)$ is a root of $u_i(X)$, the left-hand side is $t_i(x(R))$, so this is equivalent to $u_{i,P}(x(R)) = 0$, which proves correctness.

\medskip

We now analyze the complexity. The total number of iterations of the main loop is $\lvert I\rvert \cdot \lvert J\rvert$, where $\lvert J\rvert = \ell$ if $E(\F_q)[\ell^m]/E(\F_q)[\ell^{m-1}]$ is cyclic, and $\lvert J\rvert = \ell^2$ otherwise. 

In each iteration, the costs are the following:
\begin{itemize}[noitemsep,nosep]
\item computing $s_i(X)$, $\varphi_{i,x}(X)$ and $\varphi_{i,y}(X)$ costs $\Otilde(\deg(D_i))$, since each involves a constant number of multiplications and one inversion modulo $u_i(X)$;
\item the computation of $t_i(X)$ costs $\Otilde(\log q\,\deg(D_i))$, see Remark~\ref{rem:computing_psi};
\item the final GCD and remainder cost $\Otilde(\deg(D_i))$.
\end{itemize}

The dominant term is therefore $\Otilde(\log q \deg(D_i))$, and summing over all loop iterations yields the stated complexity bound.
\end{proof}

Algorithm~\ref{algo:EllipticCurveFactorizationAtLevelM} performs only one level of the decomposition: assuming that all points of $D$ lie in $E(\F_q)[N_\ell\cdot\ell^m]$, it splits $D$ according to their classes modulo $E(\F_q)[N_\ell\cdot\ell^{m-1}]$. It is not recursive, and its outputs are subdivisors of $D$: the translations by $-P$, for $P\in J$, only serve to test whether $[R-P]$ lies in $E(\F_q)[N_\ell\cdot\ell^{m-1}]$, and the translated divisors themselves are not returned. We now present Algorithm~\ref{algo:ellipticcurveprimaryfacto}, which chains these levels for $m, m-1, \ldots, 1$ and thereby separates points according to their full projection onto $E(\F_q)[\ell^\infty]$: if $P$ and $Q$ are two points of $D$ such that $[N_\ell \cdot P] \ne [N_\ell \cdot Q]$, then they end up in distinct output divisors.

The articulation between the two algorithms is as follows. At level $m$, Algorithm~\ref{algo:ellipticcurveprimaryfacto} builds the set of representatives $J$ from the generators $G_1, G_2$ (Line~\ref{line:ECPF:J}) and calls Algorithm~\ref{algo:EllipticCurveFactorizationAtLevelM} (Line~\ref{line:ECPF:levelm}). Each resulting divisor $D_P$ has its support in the coset $P + E(\F_q)[N_\ell\cdot\ell^{m-1}]$, and is therefore not a valid input for the next level in general. It is translated by $-P$ using Algorithm~\ref{algo:splitdivisortranslation}, which brings its support into $E(\F_q)[N_\ell\cdot\ell^{m-1}]$; the result is decomposed recursively at level $m-1$ (Line~\ref{line:ECPF:recur2}), and the output divisors are translated back by $P$. Moreover, Algorithm~\ref{algo:EllipticCurveFactorizationAtLevelM} requires that neither the support of $D$ nor its set of opposites meets $J$, since the translation formulas are not defined at such points. Algorithm~\ref{algo:ellipticcurveprimaryfacto} enforces this condition beforehand (Line~\ref{line:ECPF:if}): a point $P$ of $D$ such that $P\in J$ or $-P\in J$ lies in $E(\F_q)[\ell^m]$, so that its class is known in advance, and it is set aside before the recursive call at Line~\ref{line:ECPF:recur1}.

The level $m$ is an input parameter of Algorithm~\ref{algo:ellipticcurveprimaryfacto}, decremented along the recursion. In the top-level algorithm (Algorithm~\ref{algo:ellipticcurverootfinding}), we call it with the maximal value $m = \val_\ell(\exponent(E(\F_q)))$, ensuring that the required condition on $D$ is always satisfied.

\begin{algorithm}
  \caption{\EllipticCurvePrimaryFactorization{}: Partial factorization of a split divisor by projecting on $E(\F_q)[\ell^\infty]$.\label{algo:ellipticcurveprimaryfacto}}
  \INPUT{
  \begin{itemize}[algo]
      \item An elliptic curve $E$ over $\F_{q}$ given by a short Weierstrass
        equation;
      \item A prime number $\ell$ dividing $\lvert E(\F_{q})\rvert$;
      \item A nonnegative integer $m\leq \val_\ell(\exponent(E(\F_q)))$;
      \item Generators $G_1, G_2\in E(\F_q)$ of the group $E(\F_q)[\ell^\infty]$ such that $\order(G_1)\cdot\order(G_2) = \ell^{\val_\ell(\lvert E(\F_q)\rvert)}$;
      \item A split divisor $D$ on $E$, such that all points in its support lie in $E(\F_q)[N_\ell\cdot\ell^m]$;
      \item A nonsquare $\beta$ in $\F_{q}^\times$.
    \end{itemize}
  }
  \OUTPUT{
  \begin{itemize}[algo]
      \item A finite set $S\subset E(\F_q)[\ell^m]$;
      \item A finite set of tuples $\{(Q, D^{(Q)})\}_{Q\in S}$ where $D^{(Q)}$ is the split divisor formed by the points $R\in D$
      such that $[R-Q]\in E(\F_q)[N_\ell]$, and $\sum_{Q\in S} D^{(Q)} = D$.
   \end{itemize}
  }
  \lIf{$m=0$}{
    Return $\{\infty\}, \{(\infty, D)\}$\label{line:ECPF:m0}
  }
  \lIf{$D = 0$}{
    Return $\{\}, \{\}$\label{line:ECPF:n0}
  }
  $\nu_1\gets\val_\ell(\order(G_1))$, $\nu_2\gets\val_\ell(\order(G_2))$\;
  $J \gets \{\lambda_1\cdot\ell^{\nu_1-m}\cdot G_1 + \lambda_2\cdot\ell^{\nu_2-m}\cdot  G_2 : \lambda_1, \lambda_2\in\intrange 0{\ell-1}\}$\label{line:ECPF:J}\;
  \tcc{Whenever $m>\nu_i$, the term of index $i$ is omitted. Hence $\lvert J\rvert = \ell^2$ if $m\leq\min(\nu_1,\nu_2)$, and $\lvert J\rvert = \ell$ otherwise.}
  \eIf{$D$ contains a point $P$ such that $P\in J$ or $-P\in J$\label{line:ECPF:if}}{
    $\widetilde S, \{(Q, \widetilde D^{(Q)})\}_{Q\in \widetilde S}\gets \EllipticCurvePrimaryFactorization(E, \ell, m, (G_1, G_2), D-P, \beta)\label{line:ECPF:recur1}$\;
    \tcc{Since $P\in J$ or $-P\in J$, we have $P\in E(\F_q)[\ell^m]$.}
    \leIf{$P\in \widetilde S$}{
      $S\gets \widetilde S$;
      $D^{(P)} \gets \widetilde D^{(P)} + P$\;
    }{
      $S\gets\widetilde S\cup \{P\}$;
      $D^{(P)} \gets P$
    }
    Return $S, \{(Q, D^{(Q)})\}_{Q\in S}$\;
  }{
    $(D_P)_{P\in J}\gets \EllipticCurvePrimaryFactorizationAtLevelm{}(E, \ell, m, J, D)$\label{line:ECPF:levelm}\;
    $S\gets\{\}$\;
    \For{$P\in J$\label{line:ECPF:loopJ}}{
      \If{$D_P\ne 0$}{
        $\widetilde D_P \gets \SplitDivisorTranslation(E, D_P, [-P], \beta)$\;
        $S_P, \{(Q, \widetilde D_P^{(Q)})\}_{Q\in S_P}\gets \EllipticCurvePrimaryFactorization{}(E, \ell, m-1, (G_1, G_2), \widetilde D_P, \beta)$\label{line:ECPF:recur2}\;
        $S\gets S\cup\{Q + P : Q\in S_P\}$\;
        \For{$Q\in S_P$}{
          $D^{(P+Q)} \gets \SplitDivisorTranslation(E, \widetilde D_P^{(Q)}, P, \beta)$\;
        }
      }
    }
  }
  Return $S, \{(Q, D^{(Q)})\}_{Q\in S}$.
\end{algorithm}

\begin{proposition}\label{prop:ellipticcurveprimaryfacto}
  Algorithm~\ref{algo:ellipticcurveprimaryfacto} (\EllipticCurvePrimaryFactorization{}) terminates and is correct. It performs at most $\Otilde(m(\ell^2\deg(D)\log q+\deg(D)\log^2 q+\deg(D)^\omega))$ operations in $\F_q$. This complexity bound can be improved to $\Otilde(m(\ell\deg(D)\log q+\deg(D)\log^2 q+\deg(D)^\omega))$ if $E(\F_q)[\ell^m]$ is cyclic.
\end{proposition}

\begin{proof}
We begin by proving termination. The termination of the subroutines \EllipticCurvePrimaryFactorizationAtLevelm{} and \SplitDivisorTranslation{} is established in Propositions~\ref{prop:ECFactoAtLevelM} and~\ref{prop:splitdivisortranslation}. All loops in \EllipticCurvePrimaryFactorization{} are finite, so it remains to prove that the recursive calls at Lines~\ref{line:ECPF:recur1} and~\ref{line:ECPF:recur2} terminate.

This follows by induction. The base cases $m=0$ and $D=0$ terminate (Lines~\ref{line:ECPF:m0} and~\ref{line:ECPF:n0}). In the recursive calls, either $m$ decreases (Line~\ref{line:ECPF:recur2}) or $\deg(D)$ decreases (Line~\ref{line:ECPF:recur1}), so termination follows from the induction hypothesis.

\medskip

We now prove correctness. We first verify that all subroutine calls are made with valid inputs. The set $J$ computed at Line~\ref{line:ECPF:J} forms a system of representatives of $E(\F_q)[\ell^m]/E(\F_q)[\ell^{m-1}]$, so the input to \EllipticCurvePrimaryFactorizationAtLevelm{} at Line~\ref{line:ECPF:levelm} is valid. 

Next, we check that the recursive call at Line~\ref{line:ECPF:recur2} is valid, i.e.\ that all points in $\widetilde D_P$ lie in $E(\F_q)[N_\ell \cdot \ell^{m-1}]$. This follows from the correctness of \EllipticCurvePrimaryFactorizationAtLevelm{} and \SplitDivisorTranslation{} (Propositions~\ref{prop:ECFactoAtLevelM} and~\ref{prop:splitdivisortranslation}).

The two calls to \SplitDivisorTranslation{} in the loop at Line~\ref{line:ECPF:loopJ} are valid as well. For the translation of $D_P$ by $-P$, the points $\pm P$ do not lie in the support of $D_P$, by the condition at Line~\ref{line:ECPF:if}. For the translation of $\widetilde D_P^{(Q)}$ by $P$, a point of its support equal to $\pm P$ would be of the form $[R-P]$ with $R\in D_P$ and $R\in\{2P,\infty\}$. If $R = 2P$, then $P = [R-P]$ lies in $E(\F_q)[\ell^m]\cap E(\F_q)[N_\ell\cdot\ell^{m-1}] = E(\F_q)[\ell^{m-1}]$, hence $P = R = \infty$ which is excluded by the condition at Line~\ref{line:ECPF:if}, since $\infty\in J$.

We now prove that the output is correct. We proceed by induction on $m$ and $\deg(D)$. The base cases $m=0$ and $\deg(D)=0$ are handled at Lines~\ref{line:ECPF:m0} and~\ref{line:ECPF:n0}. Assume now that $m>0$ and $\deg(D)>0$.

We distinguish two cases, depending on the condition at Line~\ref{line:ECPF:if}. If the condition holds, the algorithm extracts a point $P$ and recursively processes $D-P$, which yields a correct decomposition by induction. Then $P$ is added to the result. Since $P\in J$ or $-P\in J$, and $J\subset E(\F_q)[\ell^m]$, we have $P\in E(\F_q)[\ell^m]$. As $\gcd(N_\ell,\ell)=1$, the intersection $E(\F_q)[\ell^m]\cap E(\F_q)[N_\ell]$ is trivial, so the unique $Q\in E(\F_q)[\ell^m]$ with $[P-Q]\in E(\F_q)[N_\ell]$ is $Q=P$. This is the class under which the algorithm adds $P$.

Otherwise, let $R$ be a point in $D$. The subroutine \EllipticCurvePrimaryFactorizationAtLevelm{} decomposes $D$ so that $R$ belongs to $D_P$, where $P \in J$ is such that $R - P \in E(\F_q)[N_\ell \cdot \ell^{m-1}]$. Hence $R - P \in \widetilde D_P$. By the induction hypothesis, the recursive call at Line~\ref{line:ECPF:recur2} returns a decomposition in which $R - P$ belongs to some $\widetilde D_P^{(Q)}$, with $Q \in E(\F_q)[\ell^{m-1}]$ and $R - P - Q \in E(\F_q)[N_\ell]$.
Translating back by $P$, we obtain that $R$ belongs to $D^{(P+Q)}$, and $P+Q \in S$ is an $\ell^m$-torsion point, as required.

\medskip

We now analyze the complexity. The total cost is dominated by the subroutine calls, as the additional operations are negligible in comparison. Write $d := \deg(D)$ and, for every integer $d'\geq 0$,
\[
T(d') := \ell^2 d'\log q + d'\log^2 q + d'^\omega.
\]
The function $T$ is nondecreasing and, since $\omega\geq 1$, superadditive: $T(d_1)+\cdots+T(d_r)\leq T(d_1+\cdots+d_r)$. As in the proof of Proposition~\ref{prop:twoadicfactowitheval}, we sum the costs over the tree of recursive calls, with explicit constants that do not depend on the depth.

We group the calls according to their level parameter $k\in\intrange 0m$; the calls at level $0$ perform no operation (Line~\ref{line:ECPF:m0}). For $k\geq 1$, the calls at level $k$ form disjoint chains, each made of an entry call, which is either the top-level call or a call made at Line~\ref{line:ECPF:recur2} by a call at level $k+1$, followed by successive calls at Line~\ref{line:ECPF:recur1}. Consider such a chain, whose entry call receives a divisor of degree $d'$; we exclude from its cost the entry calls at level $k-1$ that it makes at Line~\ref{line:ECPF:recur2}.

The points removed at Line~\ref{line:ECPF:recur1} are pairwise distinct elements of $J\cup(-J)$, where $J$ depends only on $k$, $G_1$ and $G_2$, so the chain makes at most $\min(d',2\ell^2)$ calls at Line~\ref{line:ECPF:recur1}. Each of them rebuilds $J$ and finds a point of the divisor whose abscissa is that of a point of $J$ by one GCD, at a cost of $\Otilde(d'+\ell^2+\log q)$, hence $\Otilde(\ell^2 d'+d'\log q)$ in total. The last call of the chain may also call \EllipticCurvePrimaryFactorizationAtLevelm{} on a divisor of degree $d''\leq d'$, at a cost of $\Otilde(\ell^2 d''\log q)$ (Proposition~\ref{prop:ECFactoAtLevelM}), followed by the translations in the loop at Line~\ref{line:ECPF:loopJ}, which cost $\Otilde(d''\log^2 q+d''^\omega)$ in total by Proposition~\ref{prop:splitdivisortranslation} and superadditivity. Since $\log q\geq 1$, there exist constants $K,\kappa>0$, independent of the inputs, such that the chain costs at most $K\,T(d')\log\bigl(\exp(1)+T(d')\bigr)^\kappa$ operations in $\F_q$.

The entry calls at level $k-1$ made by the chain receive the divisors $\widetilde D_P$, $P\in J$, whose degrees sum to $d''\leq d'$. By induction on $m-k$, the degrees of the entry calls at each level therefore sum to at most $d$. Set $\Lambda := \log(\exp(1)+T(d))$. As $T$ is nondecreasing and superadditive, each level costs at most $K\,T(d)\,\Lambda^\kappa$, and the total cost is at most $K\,m\,T(d)\,\Lambda^\kappa$. Since $\Lambda\leq\log\bigl(\exp(1)+m\,T(d)\bigr)$ for $m\geq 1$, this yields the bound $\Otilde(m\,T(d))$.

If $E(\F_q)[\ell^m]$ is cyclic, then for every $k\in\intrange 1m$, the quotient $E(\F_q)[\ell^k]/E(\F_q)[\ell^{k-1}]$ is cyclic, so that $\lvert J\rvert\leq\ell$. The same argument, with $\ell^2$ replaced by $\ell$ throughout (including in Proposition~\ref{prop:ECFactoAtLevelM}), yields the improved bound.
\end{proof}

We now present the top-level algorithm combining all the above subroutines as Algorithm~\ref{algo:ellipticcurverootfinding}. This is the online phase announced in Section~\ref{sec:intro_contrib}: its only input besides $f(X)$ is an elliptic preprocessing datum $\mathcal D$ (Definition~\ref{def:datum}), whose construction is deferred to Section~\ref{sec:precomp}. The algorithm proceeds in three stages: first, it lifts the roots of $f(X)$ to divisors on $E$ and its quadratic twist; second, it recursively decomposes these divisors using projections on $\ell$-Sylow subgroups; finally, it recovers the roots from the resulting linear factors.

Throughout, we write $E: Y^2 = h(X)$ for the curve of $\mathcal D$, with $h(X) = X^3+aX+b$. Recall from Section~\ref{sec:prelim} that the quadratic twist $\widetilde E$ of $E$ with respect to $\beta$ has Weierstrass equation $Y^2 = \widetilde h(X)$, where
\[
\widetilde h(X) = \beta^{-3}h(\beta X) = X^3 + a\beta^{-2}X + b\beta^{-3},
\]
which is the change of variables used at Lines~\ref{line:rootfinding:twistF} and~\ref{line:rootfinding:twistD} below.

\begin{remark}\label{rem:datum_suffices}
The datum $\mathcal D$ carries everything the subroutines require. The prime divisors $\ell$ of $N$ and the valuations $\val_\ell(N)$ are read off $\fac(N)$; the orders of $G_1$ and $G_2$ are computed from $\fac(N)$, so that $\exponent(E(\F_q)) = \LCM(\order(G_1),\order(G_2))$ and, for every prime $\ell$, $\val_\ell(\exponent(E(\F_q))) = \max(\val_\ell(\order(G_1)),\val_\ell(\order(G_2)))$; and since $E(\F_q)=\langle G_1\rangle\oplus\langle G_2\rangle$, the points $[N_\ell\,G_1]$ and $[N_\ell\,G_2]$ form a generating pair of the $\ell$-Sylow subgroup $E(\F_q)[\ell^\infty]$ with $\order([N_\ell\,G_1])\cdot\order([N_\ell\,G_2]) = \ell^{\val_\ell(N)}$, as required by Algorithm~\ref{algo:ellipticcurveprimaryfacto}. The same holds for $\widetilde E$.
\end{remark}

\begin{algorithm}
  \caption{\EllipticCurveRootFinding{}: Deterministic root-finding algorithm using elliptic curves.\label{algo:ellipticcurverootfinding}}
  \INPUT{
  \begin{itemize}[algo]
    \item A monic, split and separable polynomial $f \in \F_q[X]$;
    \item An elliptic preprocessing datum $\mathcal D=(\beta,E,\widetilde E, N,\widetilde N, \fac(N),\fac(\widetilde N), G_1,G_2,\widetilde G_1,\widetilde G_2)$ for $\F_q$, where $E: Y^2 = h(X)$ and $\widetilde E : Y^2 = \widetilde h(X)$ with $\widetilde h(X) = \beta^{-3}h(\beta X)$.
    \end{itemize}
  }
  \OUTPUT{
    The set of roots of $f$.
  }
  \lIf{$\GCD(f(X), h(X))\ne 1$}{
    Compute the roots of $\GCD(f(X), h(X))$ as the $x$-coordinates of the rational $2$-torsion points $(\order(G_i)/2)\cdot G_i$, for those $i\in\{1,2\}$ with $\order(G_i)$ even, and of their sum; replace $f(X)$ by $f(X)/\GCD(f(X), h(X))$\label{line:rootfinding:extraroots}
  }
  \tcc{The roots of $h(X)$ lying in $\F_q$ are exactly the $x$-coordinates of the points of order $2$ in $E(\F_q)$, and $E(\F_q)[2]$ is generated by the points above, so it suffices to keep those $x$-coordinates that are roots of $f(X)$.}
  $u_0(X), u_1(X), v_0(X), v_1(X)\gets \FiniteFieldEmbedding(q, h(X), f(X), \beta)$\;
  $F \gets \{u_0(X)\}$\;
  \For{each prime $\ell$ dividing $N$\label{line:rootfinding:loopell}}{
    \tcc{The primes $\ell$ and the valuations $\val_\ell(N)$ are read off $\fac(N)$.}
    $N_\ell\gets N /\ell^{\val_\ell(N)}$\;
    $m_\ell\gets \max(\val_\ell(\order(G_1)), \val_\ell(\order(G_2)))$\label{line:rootfinding:mell}\;
    \tcc{$m_\ell = \val_\ell(\exponent(E(\F_q)))$, see Remark~\ref{rem:datum_suffices}.}
    \For{every factor $u(X)$ in the set $F$}{
      $D \gets $ the simple divisor on $E$ with coordinates $(u(X), v_0(X)\bmod u(X))$\label{line:rootfinding:defD}\;
      $S_u, \{(Q, D_u^{(Q)})\}_{Q\in S_u}\gets \EllipticCurvePrimaryFactorization(E, \ell, m_\ell, (N_\ell\cdot G_1, N_\ell\cdot G_2), D, \beta)$\label{line:rootfinding:callprimaryfacto}\;
    }
    replace $F$ by the set of $u$-coordinates of the simple divisors occurring in all $D_u^{(Q)}$\;
  }
  $\widetilde F \gets \{\beta^{-\deg(u_1)}u_1(\beta X)\}$\label{line:rootfinding:twistF}\;
  \For{every prime $\ell$ dividing $\widetilde N$}{
    $N_\ell\gets \widetilde N /\ell^{\val_\ell(\widetilde N)}$\;
    $\widetilde m_\ell\gets \max(\val_\ell(\order(\widetilde G_1)), \val_\ell(\order(\widetilde G_2)))$\label{line:rootfinding:twistmell}\;
    \tcc{$\widetilde m_\ell = \val_\ell(\exponent(\widetilde E(\F_q)))$, see Remark~\ref{rem:datum_suffices}.}
    \For{every factor $\widetilde u(X)$ in the set $\widetilde F$}{
      $\widetilde D \gets $ the simple divisor on $\widetilde E$ with coordinates $(\widetilde u(X), v_1(\beta X)/\beta\bmod \widetilde u(X))$\label{line:rootfinding:twistD}\;
      $S_{\widetilde u}, \{(Q, \widetilde D_{\widetilde u}^{(Q)})\}_{Q\in S_{\widetilde u}}\gets \EllipticCurvePrimaryFactorization(\widetilde E, \ell, \widetilde m_\ell, (N_\ell\cdot \widetilde G_1, N_\ell\cdot \widetilde G_2), \widetilde D, \beta)$\label{line:rootfinding:callprimaryfacto2}\;
    }
    replace $\widetilde F$ by the set of $u$-coordinates of the simple divisors occurring in all $\widetilde D_{\widetilde u}^{(Q)}$\;
  }
  Return the set $\{-u(0)\}_{u(X)\in F}\cup\{-\beta\widetilde u(0)\}_{\widetilde u(X)\in\widetilde F}$.
  \tcc{If needed, also join the roots of $\GCD(f(X), h(X))$ computed at Line~\ref{line:rootfinding:extraroots}.}
\end{algorithm}

\begin{theorem}\label{thm:rootfinding}
Algorithm~\ref{algo:ellipticcurverootfinding} (\EllipticCurveRootFinding{}) terminates and is correct.  It performs at most 
$\Otilde(\Xi(\mathcal D)^2\deg(f)\log^2 q+\deg(f)^\omega\log q+\deg(f)\log^3 q)$ arithmetic operations in $\F_q$.
\end{theorem}

\begin{proof}
The termination is straightforward: all loops in the algorithm are unconditional, and the termination of the subroutines \FiniteFieldEmbedding{} and \EllipticCurvePrimaryFactorization{} has been established in Propositions~\ref{prop:finitefieldembedding} and~\ref{prop:ellipticcurveprimaryfacto}.

\medskip

We now prove correctness. We first look at Line~\ref{line:rootfinding:extraroots}. The roots of $h(X)$ lying in $\F_q$ are the $x$-coordinates of the points of $E(\F_q)[2]$. Since $E(\F_q)=\langle G_1\rangle\oplus\langle G_2\rangle$, the group $E(\F_q)[2]$ is generated by $(\order(G_i)/2)\cdot G_i$ for those $i$ with $\order(G_i)$ even, so these $x$-coordinates are obtained by two scalar multiplications. The roots of $\GCD(f(X),h(X))$ are exactly those at which $f(X)$ vanishes. After this step $f(X)$ is coprime with $h(X)$.

By Proposition~\ref{prop:finitefieldembedding}, we have $f(X)=u_0(X)\cdot u_1(X)$, so the set of roots of $f(X)$ is the union of the roots of $u_0(X)$ and $u_1(X)$. We show that the roots of $u_0(X)$ (resp.~$u_1(X)$) are precisely $\{-u(0)\}_{u\in F}$ (resp.~$\{-\beta\widetilde u(0)\}_{\widetilde u\in \widetilde F}$).

First, since \EllipticCurvePrimaryFactorization{} is correct (Proposition~\ref{prop:ellipticcurveprimaryfacto}), at every stage of the algorithm the product of all polynomials in $F$ (resp.~$\widetilde F$) equals $u_0(X)$ (resp.~$\beta^{-\deg(u_1)}u_1(\beta X)$). It remains to prove that, at the end of the algorithm, all polynomials in $F$ (resp.~$\widetilde F$) have degree $1$.

We detail the argument for $F$, the case of $\widetilde F$ being analogous. Suppose by contradiction that there exists $u(X)\in F$  of degree greater than $1$ at the end of the algorithm. Since $f(X)$ is split, $u(X)$ has two distinct roots $x,x' \in \F_q$. As $u(X)\mid u_0(X)$, both $h(x)$ and $h(x')$ are nonzero squares. Let $y=v_0(x)$ and $y'=v_0(x')$, and consider the points $P=(x,y)$ and $P'=(x',y')$ in $E(\F_q)$.

The integers $N/\ell^{\val_\ell(N)}$, for $\ell$ ranging over the primes dividing $N$, have GCD $1$. Since $P\neq P'$, the order of $[P-P']$ is not $1$, so it does not divide all of them: there exists a prime $\ell$ dividing $N$ such that
\[
\bigl[(N/\ell^{\val_\ell(N)})\cdot P\bigr] \ne \bigl[(N/\ell^{\val_\ell(N)})\cdot P'\bigr].
\]
Hence $P$ and $P'$ lie in distinct components after the call to \EllipticCurvePrimaryFactorization{} at Line~\ref{line:rootfinding:callprimaryfacto} for this value of $\ell$, contradicting the fact that they remain roots of the same polynomial in $F$ at the end of the algorithm.

\medskip

We now analyze the complexity. The dominant contributions come from the subroutines \FiniteFieldEmbedding{} and \EllipticCurvePrimaryFactorization{}, as the remaining operations are negligible.

By Proposition~\ref{prop:finitefieldembedding}, \FiniteFieldEmbedding{} requires $\Otilde(\log^2 q\deg(f))$ operations. Each call to \EllipticCurvePrimaryFactorization{} at Line~\ref{line:rootfinding:callprimaryfacto} costs
\[
\Otilde\bigl(\mu_\ell(\ell^2\deg(D)\log q + \deg(D)^\omega + \deg(D)\log^2 q)\bigr),
\]
where $\mu_\ell := \val_\ell(N)$, using that the parameter $m_\ell$ passed at Line~\ref{line:rootfinding:mell} satisfies $m_\ell\leq\mu_\ell$ (Proposition~\ref{prop:ellipticcurveprimaryfacto}).

Let $d_u$ denote the degree of the divisor $D$ defined at Line~\ref{line:rootfinding:defD}. Then $\sum_{u\in F} d_u = \deg(u_0)$. Therefore, the total cost of all calls at Line~\ref{line:rootfinding:callprimaryfacto} is bounded by
\[
\begin{aligned}
&\Otilde\left(\sum_\ell \mu_\ell \sum_{u\in F} (\ell^2 d_u\log q + d_u^\omega + d_u \log^2 q)\right) \\
&\qquad\le \Otilde\left(\sum_\ell \mu_\ell (\ell^2\deg(u_0)\log q + \deg(u_0)^\omega + \deg(u_0)\log^2 q)\right) \\
&\qquad\le \Otilde\bigl(\Xi(\mathcal D)^2\deg(u_0)\log^2 q + \deg(u_0)^\omega\log q + \deg(u_0)\log^3 q\bigr),
\end{aligned}
\]
where we used $\ell \le \Xi(\mathcal D)$ and $\sum_\ell \mu_\ell \le \log_2 N = O(\log q)$.

Similarly, the total cost of the calls at Line~\ref{line:rootfinding:callprimaryfacto2} is bounded by
\[
\Otilde\bigl(\Xi(\mathcal D)^2\deg(u_1)\log^2 q + \deg(u_1)^\omega\log q + \deg(u_1)\log^3 q\bigr).
\]

Summing all contributions yields the claimed bound.
\end{proof}

\begin{remark}\label{rem:cyclic_rootfinding}
The complexity in Theorem~\ref{thm:rootfinding} improves to
\[
\Otilde\bigl(\Xi(\mathcal D)\deg(f)\log^2 q+\deg(f)^\omega\log q+\deg(f)\log^3 q\bigr)
\]
if both $E(\F_q)$ and $\widetilde E(\F_q)$ are cyclic, provided that a generator of each group is given as input. This follows from the improved bound in Proposition~\ref{prop:ellipticcurveprimaryfacto} in the cyclic case.
\end{remark}

\section{Computing and verifying elliptic preprocessing data}\label{sec:precomp}

This section is devoted to the offline phase: constructing an elliptic preprocessing datum $\mathcal D$ for $\F_q$ with $\Xi(\mathcal D)$ small. These computations depend on $q$, but not on $f(X)$. We consider the components of $\mathcal D$ successively: the nonsquare $\beta$, the curve $E$ (Section~\ref{sec:cm_method}), the orders $N,\widetilde N$ and their factorizations, and the bases (Section~\ref{sec:precom_gens}). We conclude in Section~\ref{sec:verification} by showing that a preprocessing datum can be \emph{verified} much more easily than it can be constructed.

When $q=p$ is prime, all components but the curve are within reach of deterministic polynomial-time methods under GRH. For $e>1$, this holds for all of them except the bases (see Section~\ref{sec:precom_gens}). The curve is the main obstruction, and this question is isolated in Problem~\ref{prob:precomp}.

We begin with the nonsquare $\beta$. For $a \in \F_p^\times$, we have $a^{(q-1)/2} = \big(a^{(p-1)/2}\big)^{e}$, so a nonsquare of $\F_p$ remains a nonsquare in $\F_q$ if and only if $e$ is odd. In that case, under GRH, \cite{ankeny1952least} guarantees a quadratic non-residue modulo $p$ smaller than $C(\log p)^2$ for some constant $C>0$, which is found deterministically in polynomial time by testing successive integers. 

If $e$ is even, every element of $\F_p$ is a square in $\F_q$, but a nonsquare of $\F_q$ can then be computed deterministically and unconditionally in polynomial time~\cite[Cor.~1.3]{bhargava2017irreducibility}. For completeness, we give an explicit construction. Set $Q := p^{e/2} = \sqrt q$, and pick $\theta\in\F_q\setminus\F_Q$, for instance an element of the given $\F_p$-basis of $\F_q$ such that $\theta^Q\neq\theta$. The element $\delta := \theta-\theta^Q$ satisfies $\delta^Q = -\delta$, hence
\[
\delta^{(q-1)/2} = \bigl(\delta^{Q-1}\bigr)^{(Q+1)/2} = (-1)^{(Q+1)/2}.
\]
If $Q\equiv 1 \bmod 4$, then $\delta$ is a nonsquare; this is the case $r=2$ of~\cite[Thm.~3.1]{bhargava2017irreducibility}.

Otherwise, $p\equiv 3 \bmod 4$ and $e/2$ is odd. Then $\delta^{2Q} = \delta^2$ so $\delta^2$ lies in $\F_Q$ but is not a square there, and $-1$ is a nonsquare in $\F_Q$ since $Q\equiv 3\bmod 4$; hence $-\delta^2$ is a square in $\F_Q$, and $s := (-\delta^2)^{(Q+1)/4}$ satisfies $s^2 = -\delta^2$. The element $\eta := \delta/s$ thus satisfies $\eta^2=-1$; in particular $\eta\in\F_{p^2}$ and $\eta^p = -\eta$. Since $\bigl(\frac{1}{p}\bigr) = 1$ and $\bigl(\frac{p-1}{p}\bigr) = -1$, a dichotomy on $\intrange{1}{p-1}$ finds, with $O(\log p)$ evaluations of the Legendre symbol, an integer $k$ such that $\bigl(\frac{k}{p}\bigr) = 1$ and $\bigl(\frac{k+1}{p}\bigr) = -1$. Then $\nu := k^{(p+1)/4}\in\F_p$ satisfies $\nu^2 = k$, and $\beta := \nu+\eta$ is a nonsquare of $\F_q$. Indeed, $\beta^{p+1} = (\nu+\eta)(\nu-\eta) = k+1$, so that
\[
\beta^{(q-1)/2} = \Bigl(\bigl(\beta^{p+1}\bigr)^{(p-1)/2}\Bigr)^{(q-1)/(p^2-1)} = (-1)^{(q-1)/(p^2-1)} = -1,
\]
since $(q-1)/(p^2-1)$ is odd.

For even $e$, in both cases, the computation is deterministic, unconditional, and runs in time polynomial in $\log q$.

\subsection{Constructing the curve: the CM method and its limitations}\label{sec:cm_method}

Since $\Xi(E)=\Xi(\widetilde E)$, searching for a suitable curve amounts to searching over pairs $\{E,\widetilde E\}$ of quadratically twisted curves.

When \( 4p = U^2 + dV^2 \) with \( p \) prime and \( d \) small, a general method for constructing elliptic curves over prime fields with prescribed Frobenius trace is the Complex Multiplication (CM) method. We refer to~\cite{morain1991building,atkin1993elliptic} and~\cite[Sec.~18]{cohen2005handbook} for background on this method. It relies on algebraic number theory in the imaginary quadratic field \( \mathbb{Q}(\sqrt{-d}) \). In our setting, we seek values of $U$ such that both $q+1-U$ and $q+1+U$ have no large prime factors.

Given a trace \( U \in \intrange{0}{2\sqrt{p}} \) allowed by the Hasse–Weil bound and not divisible by \( p \), the CM method proceeds by first constructing the Hilbert class polynomial \( H_d(X) \) associated with \( \mathbb{Q}(\sqrt{-d}) \), where $d$ is the squarefree integer such that $d C^2 = 4p-U^2$ for $C\in\Z$. This can be done deterministically in time polynomial in \( \log p \) and \( d \); see~\cite{atkin1993elliptic} and~\cite[Sec.~4]{morain1991building}. 

An elliptic curve with Frobenius trace \( U \) can then be obtained from a root of \( H_d(X) \bmod p \). By Dirichlet's class number formula, the degree of \( H_d(X) \), which equals the class number of \( \mathbb{Q}(\sqrt{-d}) \), is \( O(\sqrt{d}\log d) \). Therefore, when \( d \) is sufficiently small, a root of \( H_d(X) \bmod p \) can be found efficiently using a probabilistic algorithm.

The Hilbert class polynomial also enjoys strong Galois-theoretic properties: in particular, its splitting field is an abelian extension of \( \mathbb{Q}(\sqrt{-d}) \). Moreover, this splitting field has degree at most \( 2\deg(H_d) \) over \( \mathbb{Q} \). It follows from~\cite[Cor.~5.6]{ronyai1992galois} that, under GRH, a root of \( H_d(X) \bmod p \) can be found deterministically in time polynomial in \( \log p \), \( \deg(H_d) \) and in the bit-size of the coefficients of $H_d(X)$.

Unfortunately, the CM method is likely not sufficient to construct elliptic curves $E$ with sufficiently small $\Xi(E)$. Indeed, the requirement that $d$ be small is quite restrictive, and it is therefore unlikely that, for every $p$, there exists an elliptic curve such that both $d$ and $\Xi(E)$ are polynomial in $\log(p)$.

\subsection{Distribution of elliptic curves with smooth group orders}\label{sec:distrib_smooth}

This subsection relates Conjecture~\ref{conj:main_conj} to classical questions on the additive behaviour of smooth numbers, and explains what the standard heuristics predict. We refer to~\cite{hildebrand1993integers} for a survey of the distribution of smooth numbers.

\paragraph{Relation to additive problems on smooth numbers.}
Writing $N := q+1-U$ and $\widetilde N := q+1+U$, we have $N+\widetilde N = 2(q+1)$. Conjecture~\ref{conj:main_conj} therefore says that $2(q+1)$ can be written as a sum of two $(\log q)^{O(1)}$-smooth integers, subject to the constraint that the two summands be within $2\sqrt q$ of $q+1$.

Without that constraint, this is a question of Erd\H os and Graham~\cite[p.~70]{erdos1980old}, who asked whether every large $n$ is a sum of two $n^{\varepsilon}$-smooth integers. The best known result, due to Balog~\cite{balog1989additive}, provides $n^{4/(9\sqrt{\exp(1)})+\varepsilon}$-smooth summands, where $4/(9\sqrt{\exp(1)})\approx 0.27$. Even whether both summands can always be taken $n^{o(1)}$-smooth is open, while the equidistribution heuristic discussed below predicts that a power of $\log n$ suffices.

We are interested in smooth numbers in the Hasse--Weil interval, which is a short interval of length $4\sqrt q$ around $q+1$. The relevance of that interval comes from the following theorem:

\begin{theorem}\cite[Thm.~4.1]{waterhouse1969abelian}
  For any $N$ in the Hasse--Weil interval $\intrange{q+1-2\sqrt q}{q+1+2\sqrt q}$ such that $q+1-N$ is coprime to $q$, there exists an elliptic curve defined over $\F_q$ with $\lvert E(\F_q)\rvert = N$.
\end{theorem}

Since elliptic curves with smooth group order are useful in various contexts (e.g.~\cite{maurer1994towards,ECFFT1}), conjectures have been proposed regarding the distribution of smooth numbers in the Hasse–Weil interval.

\begin{conjecture}[Smoothness assumption]\cite[Def.~2 and Conjecture]{maurer1994towards}, \cite[Sec.~4.1.1]{maurer1999relationship}\label{conj:maurer}
The largest prime factor of the smoothest number in the Hasse--Weil interval $\intrange{q+1-2\sqrt q}{q+1+2\sqrt q}$ is at most $(\log q)^{O(1)}$.
\end{conjecture}

Conjecture~\ref{conj:maurer} is not sufficient for our purposes, since we require the quadratic twist to have a smooth order as well. Conjecture~\ref{conj:main_conj} is the corresponding statement for the pair $(N,\widetilde N)$.

\paragraph{What the standard heuristic predicts.}
Both Conjecture~\ref{conj:maurer} and Conjecture~\ref{conj:main_conj} rest on the heuristic that smooth numbers are as well distributed in the Hasse--Weil interval as they are on average. Matomäki and Radziwiłł~\cite{matomaki2016multiplicative} proved that, for every $\varepsilon>0$, there exists $C(\varepsilon)>0$ such that, for $x$ large enough, the interval $\intrange{x}{x+C(\varepsilon)\sqrt x}$ contains an $x^\varepsilon$-smooth integer. This is the scale of the Hasse--Weil interval, of length $4\sqrt q$, up to the constant $C(\varepsilon)$. Even ignoring that constant, this only provides, for every fixed $\varepsilon>0$, a $q^{\varepsilon}$-smooth group order, whereas Conjecture~\ref{conj:maurer} asks for a $(\log q)^{O(1)}$-smooth one; and it says nothing about the simultaneous smoothness of $N$ and $\widetilde N$ required by Conjecture~\ref{conj:main_conj}.

We can predict the exponent by assuming an equidistribution heuristic. Let $y := (\log q)^\alpha$ with $\alpha>0$. By the Dickman--de Bruijn estimate, an integer near $q$ is $y$-smooth with probability $\rho(\log q/\log y) = q^{-1/\alpha+o(1)}$, where $\rho$ is the Dickman function. The $4\sqrt q$ integers of the Hasse--Weil interval should thus contain $q^{1/2-1/\alpha+o(1)}$ numbers that are $y$-smooth, a quantity that grows when $\alpha>2$: this is the threshold underlying Conjecture~\ref{conj:maurer}. Treating the smoothness of $N$ and of $\widetilde N$ as independent, the $2\sqrt q$ pairs $(N,\widetilde N)$ should contain $q^{1/2-2/\alpha+o(1)}$ admissible ones, which grows when $\alpha>4$. As this exponent affects none of our complexity bounds, Conjecture~\ref{conj:main_conj} leaves the polynomial bound unspecified.

\paragraph{A cryptographic analogue.}
The same simultaneous-smoothness condition arises in isogeny-based cryptography, where one seeks primes $p$ such that $p-1$ and $p+1$ are both smooth. Considerable effort has gone into such searches~\cite{costello2021sieving}, and the smoothness bounds they reach are worse than what a single condition would allow, which is consistent with the gap between the two thresholds above.

By Corollary~\ref{cor:datum_exists}, Conjecture~\ref{conj:main_conj} guarantees the existence of an elliptic preprocessing datum $\mathcal D$ with $\Xi(\mathcal D)\leq(\log q)^{O(1)}$, for which the online phase of Algorithm~\ref{algo:ellipticcurverootfinding} runs in time polynomial in $\log q$ and $\deg(f)$. However, the conjecture is only an existence statement: it says nothing about how to construct such a datum.

\subsection{Computing the group orders, their factorizations, and the bases}\label{sec:precom_gens}

Computing the order $N = \lvert E(\F_q)\rvert$ can be done deterministically in polynomial-time by using Schoof's algorithm~\cite{schoof1985elliptic} or its heuristic improvements of Elkies and Atkin. The order of the twist is obtained via the formula $\widetilde N = 2(q+1)-N$.

The datum also records the factorizations $\fac(N)$ and $\fac(\widetilde N)$. Factoring an integer of size $q$ is of course not known to be feasible in polynomial time in general, but in our case it is: a curve is only of interest to us when $N\cdot\widetilde N$ is $\Xi(\mathcal D)$-smooth, so trial division by the integers $d = 2,3,\ldots,\Xi(\mathcal D)$ recovers both factorizations, in $\Otilde(\Xi(\mathcal D)\log^2 q)$ bit operations. Running this on a curve which turns out not to be smooth simply rejects it. This is why the factorizations belong to the datum.

In addition, Algorithm~\ref{algo:ellipticcurverootfinding} requires bases of the groups of rational points of \( E \) and \( \widetilde E \).
In general, without any smoothness assumption, the best deterministic method requires time $O(q^{1/2+\varepsilon})$~\cite{kohel2000exponential}.
However, when $q=p$ is prime and $\lvert E(\F_p)\rvert$ is smooth, one can do better. We restrict to prime fields because the results of~\cite{ronyai1992galois} used below are stated over $\F_p$ only.

\begin{proposition}
Let $p$ be a prime and $E$ an elliptic curve over $\F_p$. Under GRH, there is a deterministic algorithm to compute generators of $E(\F_p)$ whose complexity is polynomial in $\log p$ and in the largest prime factor of $\lvert E(\F_p)\rvert$.
\end{proposition}

\begin{proof}
We compute for each prime $\ell$ dividing $\lvert E(\F_p)\rvert$ a generating pair $(G_1^{(\ell)},G_2^{(\ell)})$ of the $\ell$-Sylow subgroup of $E(\F_p)$. Since $E(\F_p)$ is the direct sum of its Sylow subgroups, the points $G_1 := \sum_\ell G_1^{(\ell)}$ and $G_2 := \sum_\ell G_2^{(\ell)}$ generate $E(\F_p)$.

Fix such a prime $\ell$ and set $\mu = \val_\ell(\lvert E(\F_p)\rvert)$. We compute a generating pair of $E(\F_p)[\ell^k]$ by induction on $k$. For $k=1$, the group $E(\F_p)[\ell]$ is read off the roots lying in $\F_p$ of the division polynomial $\psi_\ell(X,Y)$ which involves only $X$ when $\ell$ is odd (resp.\ of $X^3+aX+b$ when $\ell=2$), which are computed deterministically in polynomial time under GRH by~\cite[Thm.~6.3]{ronyai1992galois}. When $\ell$ is odd, such a root $x$ yields points of $E(\F_p)$ if and only if $x^3+ax+b$ is a square in $\F_p$, and their ordinates are then computed by the Tonelli--Shanks algorithm, which is deterministic given the nonsquare $\beta$. The group $E(\F_p)[\ell]$ is isomorphic to $\Z/\ell\Z$ or $(\Z/\ell\Z)^2$, so a generating pair is obtained by picking a nonzero point $G_1$, and a point $G_2\notin\langle G_1\rangle$ if there is one, $G_2=\infty$ otherwise.

Assume now that $G_1, G_2$ generate $E(\F_p)[\ell^k]$, with $1\leq k<\mu$. Deciding whether a point of $E(\F_p)$ lies in $\ell E(\F_p)$, and computing a preimage of the multiplication by $\ell$, is again deterministic polynomial time under GRH by~\cite[Thm.~6.4]{ronyai1992galois}. Since $[\ell G_1]$ and $[\ell G_2]$ lie in $\ell E(\F_p)$, whether $[c_1G_1+c_2G_2]$ lies in $\ell E(\F_p)$ only depends on $(c_1,c_2)\bmod\ell$, and the corresponding classes form a subspace $K$ of $\F_\ell^2$, which we compute by testing the $\ell^2$ classes. Replacing $(G_1,G_2)$ by $([aG_1+bG_2],[cG_1+dG_2])$ for suitable integers $a,b,c,d$ with $ad-bc\not\equiv 0\bmod\ell$, which does not change the generated group, we may assume that $K$ is spanned by a subset of the canonical basis of $\F_\ell^2$. We then replace by a preimage under $[\ell]$ each $G_i$ lying in $\ell E(\F_p)$, and keep the others. The new pair generates $E(\F_p)[\ell^{k+1}]$: the group it generates contains the kernel $E(\F_p)[\ell]$ of $[\ell]$, and its image under $[\ell]$ is $E(\F_p)[\ell^k]\cap\ell E(\F_p) = [\ell]\bigl(E(\F_p)[\ell^{k+1}]\bigr)$.
Iterating until $k=\mu$ yields a generating pair of the $\ell$-Sylow subgroup $E(\F_p)[\ell^\mu]$, after $\mu = O(\log p)$ steps of $\ell^2$ tests each.
\end{proof}

A generating pair produced this way need not be a basis in the sense of Definition~\ref{def:datum}. We now explain how to turn a generating pair $G_1, G_2$ of $E(\F_q)$ into a basis, i.e.\ so that $\order(G_1)\cdot\order(G_2) = \lvert E(\F_q)\rvert$, which is equivalent to requiring $E(\F_q)=\langle G_1\rangle\oplus\langle G_2\rangle$. We work on each Sylow subgroup separately, so that the modifications made for one prime do not affect the others.

Fix a prime $\ell$ dividing $N$, and let $N_\ell := N/\ell^{\val_\ell(N)}$. The points $P_1 := [N_\ell\, G_1]$ and $P_2 := [N_\ell\, G_2]$ generate $E(\F_q)[\ell^\infty]$. Up to swapping them, assume that $m_1\leq m_2$, where $m_i := \val_\ell(\order(P_i))$. If $m_1 = 0$, then $P_1 = \infty$ and $E(\F_q)[\ell^\infty] = \langle P_2\rangle$. Otherwise, if for some $\lambda \in \intrange{1}{\ell-1}$ the point
\[
P_1' := [P_1 - \lambda \ell^{m_2 - m_1} P_2]
\]
satisfies $\val_\ell(\order(P_1')) < m_1$, we replace $P_1$ by $P_1'$, which does not change the group generated, and repeat. When no such $\lambda$ exists, $\langle P_1\rangle\cap\langle P_2\rangle$ is trivial. Otherwise, the unique subgroups of order $\ell$ of $\langle P_1\rangle$ and $\langle P_2\rangle$ would coincide, i.e.\ $[\ell^{m_1-1}P_1] = [\lambda\ell^{m_2-1}P_2]$ for some $\lambda\in\intrange{1}{\ell-1}$, and then $[\ell^{m_1-1}P_1'] = \infty$. Hence $E(\F_q)[\ell^\infty] = \langle P_1\rangle\oplus\langle P_2\rangle$. We order the pair so that $\order(P_1)\geq\order(P_2)$, and denote it by $P_1^{(\ell)}, P_2^{(\ell)}$.

Finally, we set $G_1 := \sum_\ell P_1^{(\ell)}$ and $G_2 := \sum_\ell P_2^{(\ell)}$. Since $E(\F_q)$ is the direct sum of its Sylow subgroups, $\langle G_i\rangle = \bigoplus_\ell \langle P_i^{(\ell)}\rangle$, so that $E(\F_q)=\langle G_1\rangle\oplus\langle G_2\rangle$. Moreover, $G_2=\infty$ when $E(\F_q)$ is cyclic, since every $E(\F_q)[\ell^\infty]$ is then cyclic and $P_2^{(\ell)} = \infty$.

For each \( \ell \), this procedure requires two scalar multiplications by $N_\ell$ and at most $\ell \log q$ computations of an \( \ell \)-adic valuation, for a total of \( O(\ell \log^2 q) \) operations in \( \mathbb{F}_q \). Summing over all primes $\ell \leq \Xi(\mathcal D)$, we obtain an overall complexity of $O(\Xi(\mathcal D)^2 \log^2 q)$ arithmetic operations in $\F_q$. A similar method is proposed at the end of the proof of~\cite[Thm.~3]{kohel2000exponential}.

Generators can also be computed probabilistically in polynomial time using the Weil pairing, following the method of~\cite[Thm.~5]{miller2004weil}, provided that the factorization of \( \GCD(N, q-1) \) is known, which is again available by trial division when $N$ is smooth.

\subsection{Verifying a preprocessing datum}\label{sec:verification}

The constructions described so far are either conditional on GRH, or, in the case of the curve itself, not known to be feasible in deterministic polynomial time at all (see Problem~\ref{prob:precomp}). Verification, by contrast, is deterministic and unconditional:

\begin{proposition}\label{prop:verification}
Given $q$ and a tuple $\mathcal D = (\beta,E,\widetilde E, N,\widetilde N,
\fac(N),\fac(\widetilde N),
G_1,G_2,\widetilde G_1,\widetilde G_2)$, checking that $\mathcal D$ is a valid elliptic preprocessing datum for $\F_q$ can be achieved deterministically in $\Otilde(\log^3 q+\Xi(\mathcal D)\log^2 q)$ bit operations.
\end{proposition}

\begin{proof}
We describe the checks on $E$; those on $\widetilde E$ are identical. Checking that $N$ lies in the Hasse--Weil interval $\intrange{q+1-2\sqrt q}{q+1+2\sqrt q}$ and that $\widetilde N = 2(q+1)-N$, that $\beta^{(q-1)/2}=-1$, that the equation $Y^2=X^3+aX+b$ of $E$ satisfies $4a^3+27b^2\neq 0$, that $\widetilde E$ has equation $Y^2=X^3+\beta^{-2}aX+\beta^{-3}b$, and that $G_1,G_2$ lie on $E$ costs $\Otilde(\log^2 q)$ bit operations.

We then check that the product of the factors listed in $\fac(N)$ equals $N$. Every prime factor of each of them then divides $N$, hence is at most $\Xi(\mathcal D)$, so trial division by $d=2,3,4,\ldots$ on all factors, decides their primality in $\Otilde(\Xi(\mathcal D)\log^2 q)$ bit operations.

Next, after checking that $[N\,G_1]=[N\,G_2]=\infty$, we compute the orders $n_1,n_2$ of $G_1,G_2$ from $\fac(N)$ in $\Otilde(\log^3 q)$ bit operations. A nontrivial intersection of $\langle G_1\rangle$ and $\langle G_2\rangle$ would contain a point of prime order $\ell\mid\gcd(n_1,n_2)$, and $\langle G_i\rangle$ has a unique subgroup of order $\ell$, generated by $[(n_i/\ell)\,G_i]$. Independence thus amounts to checking, for each such $\ell$, that $[(n_1/\ell)\,G_1]$ is none of the points $[(i n_2/\ell)\,G_2]$ for $1\leq i<\ell$, for a total of $\Otilde(\Xi(\mathcal D)\log^2 q)$ bit operations.

Finally, we check that $n_1 n_2=N$. The subgroup $\langle G_1\rangle\oplus\langle G_2\rangle$ then has order $N$, so $N$ divides $\lvert E(\F_q)\rvert$. For $q\geq 37$, we have $q+1>6\sqrt q$, so $N$ is the only multiple of $N$ in the Hasse--Weil interval, whence $\lvert E(\F_q)\rvert=N$; the fields with $q<37$ are handled by finite enumeration.
\end{proof}

\section{Experiments}\label{sec:expe}

\subsection{Experimental evidence for Conjecture~\ref{conj:main_conj}}\label{sec:expe_conj}

\begin{figure}
\centering\begin{tikzpicture}
  \begin{axis}[
title={$q$ is prime},
xlabel={$q$},
ylabel={$\min_U(\log\xi_q(U)/\log\log q)$},
ymin=0,
ymax=2,
legend style={at={(0.97,0.03)},anchor=south east,font=\scriptsize,fill opacity=0.8,text opacity=1},
]
\addplot[only marks,mark options={scale=0.2}] table {data_conjecture.data};
\addlegendentry{observed}
\addplot[no markers,thick,red] table {data_heuristic.data};
\addlegendentry{predicted threshold}
\end{axis}
\end{tikzpicture}\quad\quad
\begin{tikzpicture}
  \begin{axis}[
title={$q = p^2$ is a square},
xlabel={$q$},
ylabel={$\min_U(\log\xi_q(U)/\log\log q)$},
ymin=0,
ymax=2,
legend style={at={(0.97,0.03)},anchor=south east,font=\scriptsize,fill opacity=0.8,text opacity=1},
]
\addplot[only marks,mark options={scale=0.2}] table {data_conjecture2.data};
\addlegendentry{observed}
\addplot[no markers,thick,red] table {data_heuristic2.data};
\addlegendentry{predicted threshold}
\end{axis}
\end{tikzpicture}

\caption{Observed smoothness exponent of the best pair $(N,\widetilde N)$ in the Hasse--Weil interval, compared with the threshold predicted by the heuristic of Section~\ref{sec:distrib_smooth}.\label{fig:testing_conj}}
\end{figure}
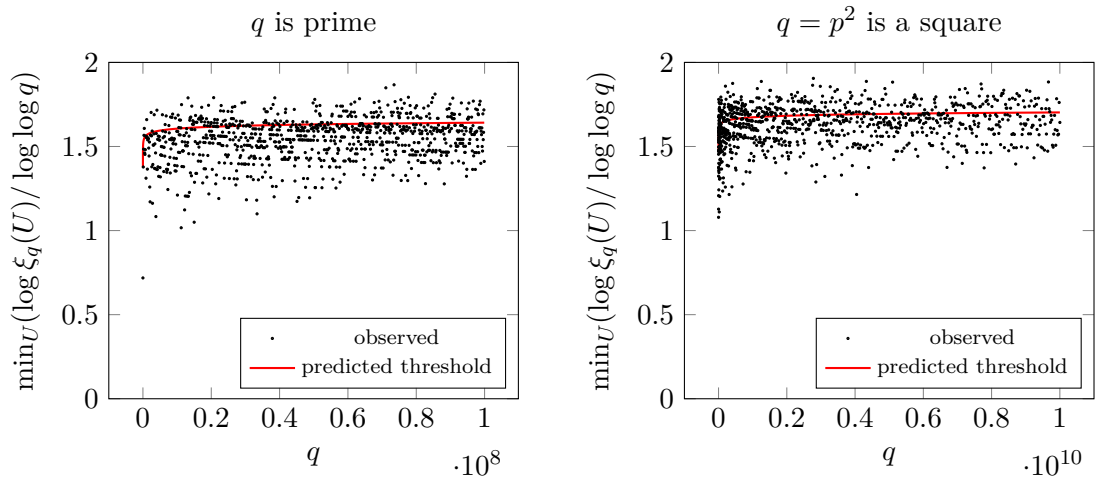

Figure~\ref{fig:testing_conj} compares the smoothness actually achievable in the Hasse--Weil interval with what the heuristic of Section~\ref{sec:distrib_smooth} predicts. For a sampled prime power $q$ and an integer $U$, we define $\xi_q(U)$ as the largest prime factor of $(q+1-U)(q+1+U)$. For each fixed $q$, we compute the minimal value of $\xi_q(U)$ over all traces $U\in\intrange{0}{2\sqrt q}$ allowed by the Hasse--Weil bound and coprime with $q$.

On the vertical axis, we plot the value
\[
\alpha := \min_U \frac{\log \xi_q(U)}{\log\log q},
\]
so that $\min_U \xi_q(U) = (\log q)^\alpha$. Over the whole range tested, and for prime fields as well as for quadratic extensions, $\alpha$ stays within a narrow band and grows slowly with $q$.

On these data points, we superpose the threshold predicted by the heuristic. For each $q$, the curve is the smallest $\alpha$ for which the expected number of admissible traces computed in Section~\ref{sec:distrib_smooth} reaches one, that is, the smallest $\alpha$ such that
\[
2\sqrt q\;\rho\!\left(\frac{\log q}{\log\bigl((\log q)^\alpha\bigr)}\right)^{2}\geq 1,
\]
where $\rho$ is the Dickman function, obtained by solving the differential equation $u\rho'(u)=-\rho(u-1)$ with $\rho(u)=1$ for $0\leq u\leq 1$. Over the sampled ranges it rises from $1.38$ to $1.64$ for prime $q$, and from $1.47$ to $1.70$ for $q=p^2$.

The observed exponent follows the predicted threshold closely, lying slightly below it on average. A small discrepancy is what one should expect, since the criterion ``expected count at least one'' ignores both the fluctuations of that count and the $o(1)$ in the Dickman estimate.

This is evidence that the heuristic of Section~\ref{sec:distrib_smooth} is accurate in the range where the minimum can be computed, which is in turn what makes the polynomial bound asserted by Conjecture~\ref{conj:main_conj} plausible. Yet, the accessible range of $q$ is too small to claim an estimate about the exponent $c$.

The Magma scripts used to generate the data points and the predicted curves are available in the files \texttt{Testing\_conjecture.mgm} and \texttt{Heuristic\_threshold.mgm} in~\cite{implementation}.

\subsection{Proof-of-concept implementation in Magma}

As a proof of concept, we implemented our algorithm in the Magma computer algebra system, and report here on experimental results obtained with this prototype. 
The implementation is available at~\cite{implementation}.
All experiments were performed on a standard laptop with an Intel Core Ultra 7 255U processor, using Magma~V2.29-6~\cite{MR1484478}.

To demonstrate the practicality of the method, we select a general prime~$p$. To rule out the use of cyclotomic methods, we choose the first safe prime (i.e., such that $(p-1)/2$ is also prime, ensuring that $\F_p^\times$ contains a large prime-order subgroup) in the interval $\intrange{2^{63}}{2^{64}}$, namely
\[
p = 9223372036854778487.
\]

The next step is the offline phase, that is, building an elliptic preprocessing datum $\mathcal D$ for $\F_p$ with $\Xi(\mathcal D)$ small. To this end, we scan $j$-invariants $j\ne 0,1728$ sequentially; for each of them, we consider the curve
\[
E_j : Y^2 = X^3 - \frac{j}{48(j-1728)}X + \frac{j}{864(j-1728)},
\]
which has $j$-invariant $j$, together with its quadratic twist, and use Magma's implementation of the SEA algorithm to compute the Frobenius trace. After a few hours of computation, the best curve we found has $j$-invariant $17698758$.

The curve $E_j$ has
\[
9223372041035794551 = 3^3\cdot 7\cdot 13\cdot 19\cdot 43\cdot 59\cdot 61\cdot 71\cdot 113\cdot 227\cdot 701
\]
rational points, while its quadratic twist has
\[
9223372032673762425 = 3\cdot 5^2\cdot 7\cdot 11\cdot 19\cdot 83\cdot 107\cdot 1549\cdot 2269\cdot 2693
\]
rational points. Therefore,
$\Xi(\mathcal D) = 2693 \approx \log(p)^{2.091}$,
which controls the complexity of the online phase.
In our implementation, the nonsquare $\beta$ is a primitive element of $\F_p$ returned by Magma, and the two bases are obtained with Magma's \texttt{Generators} function rather than with the procedure of Section~\ref{sec:precom_gens}. The resulting datum $\mathcal D$ is then certified by Proposition~\ref{prop:verification}, in negligible time.

This search explores only a tiny fraction of the possible $j$-invariants: we tested a few million values, out of approximately $2^{63}$.

\medskip

Next, we generate a random split separable polynomial $f(X)$ of degree $3000$ over $\F_p$ by sampling $3000$ distinct elements $x_1,\ldots,x_{3000}$ uniformly at random in $\F_p$, and setting $f(X) = \prod_{i=1}^{3000} (X - x_i)$.

Our (unoptimized) Magma prototype computes all roots of $f$ in about 3 minutes, the datum $\mathcal D$ being given. This experiment shows that Algorithm~\ref{algo:ellipticcurverootfinding} can be run in practice.

For comparison, the classical Mignotte--Schnorr algorithm (Appendix~\ref{sec:app_MS}) is out of reach in this setting. Its complexity is governed by the largest prime factor $\ell$ of $p-1$, which here is $(p-1)/2\approx 2^{62}$, so it would require $\ell\approx 2^{62}$ GCD computations.

\section{Perspectives and open questions}\label{sec:conclusion}

\subsection{Improving the complexity bound}\label{sec:improving_complexity}

There are several possible directions for improving the complexity bound in Theorem~\ref{thm:rootfinding}. The first one concerns the exponent $\omega$. Inspecting the proofs of Propositions~\ref{prop:splitdivisortranslation} and~\ref{prop:ECFactoAtLevelM}, only two steps cost $\deg(f)^\omega$, both in Algorithm~\ref{algo:splitdivisortranslation}: the implicit interpolation of Remark~\ref{rem:interpolation}, and the characteristic polynomial of Remark~\ref{rem:charpol}. Both are currently solved through an explicit $\deg(f)\times\deg(f)$ matrix, whose construction alone costs $\Otilde(\deg(f)^2)$ operations. 

We expect that both steps can be performed without this matrix, using power projections~\cite{shoup1994fast,bostan2003fast}. The interpolation problem of Remark~\ref{rem:interpolation} is an instance of \emph{inverse modular composition}. When the quotient ring is a field, Shoup~\cite[Thm.~3.5]{shoup1994fast} solves this problem deterministically within the cost of Brent and Kung's modular composition algorithm: he computes power projections by transposition of that algorithm, and then solves a Hankel system; see also~\cite[Sec.~1.4.2]{neiger2024faster}. In our setting, $u(X)$ is split, so that $\F_q[X]/(u(X))$ is not a field, but we may use the trace form instead of Shoup's linear form. The characteristic polynomial of Remark~\ref{rem:charpol} could be handled similarly: Shoup also computes minimal polynomials from power projections~\cite[Thm.~3.4]{shoup1994fast}. If this approach is confirmed, the exponent $\omega$ in Theorem~\ref{thm:rootfinding} would be replaced by $(\omega+1)/2$.

Going further, Neiger, Salvy, Schost and Villard~\cite{neiger2024faster} recently obtained the first algorithm for modular composition with an exponent below $3/2$, namely $O(n^{1.43})$ operations, and their techniques also apply to the computation of minimal and characteristic polynomials of elements of quotient rings such as $\F_q[X]/(u(X))$~\cite[Sec.~10.1]{neiger2024faster}. Their algorithms are randomized, and adapting them to our deterministic setting is an open question.

\subsection{Using more general curves}
\label{sec:hyperelliptic_curves}

Our approach can be extended to hyperelliptic curves of genus $g>1$. One could start by choosing a squarefree polynomial $h(X)$ of odd degree, defining a hyperelliptic curve $\mathcal H$ with equation $Y^2 = h(X)$. Algorithms~\ref{algo:TwoadicFactoWithEval} and~\ref{algo:curveembedding} do not impose restrictions on the degree of $h$, and can therefore be used to construct split divisors on $\mathcal H$ and on its quadratic twist that encode the roots of the input polynomial $f(X)$.

The next step would be to use rational points on the Jacobian of $\mathcal H$ and its quadratic twist. In this setting, division polynomials in Algorithm~\ref{algo:EllipticCurveFactorizationAtLevelM} can be replaced by Cantor's analogues for hyperelliptic curves~\cite{cantor1994analogue}.

The potential advantage of this approach is the increased flexibility in constructing groups over $\F_q$ with smooth order. However, a drawback is that the Jacobian of a genus-$g$ hyperelliptic curve has approximately $q^g$ points, making smoothness less likely. Nevertheless, CM methods could potentially be adapted to enforce smooth group orders in this setting, for instance using techniques from~\cite{balakrishnan2016constructing}.

\subsection{Using other algebraic groups}

In this paper, we focused on elliptic curves. As discussed in Section~\ref{sec:hyperelliptic_curves}, the approach may extend to more general curves. More broadly, it may be possible to use other commutative algebraic groups defined over $\F_q$.

One potential direction is to consider algebraic tori~\cite{voskresenskii2011algebraic, rubin2003torus}, whose group orders are given by evaluations of cyclotomic polynomials at $q$. While deterministic polynomial-time algorithms are known when such values are smooth~\cite{bach2001factoring}, these results are mainly of theoretical interest. Adapting the Mignotte--Schnorr framework in this setting could lead to more explicit and practical complexity bounds.

\subsection{Polynomial-time deterministic precomputation}
\label{sec:redEC}

We close by returning to Problem~\ref{prob:precomp}, an effective version of Conjecture~\ref{conj:main_conj}. For prime $q$, it amounts under GRH to finding the curve (Section~\ref{sec:precomp}).

By Proposition~\ref{prop:datum_existence}, it suffices to find a Frobenius trace $U$ such that $(q+1-U)(q+1+U)$ is $(\log q)^{O(1)}$-smooth, together with a curve realizing it. Among the roughly $4\sqrt q$ admissible traces, candidates are cheap to reject and, by Proposition~\ref{prop:verification}, cheap to certify; what is missing is a deterministic way of finding one. The CM method (Section~\ref{sec:cm_method}) only reaches traces with small discriminant, exhaustive scanning of $j$-invariants (as in our implementation) is not polynomial, and Conjecture~\ref{conj:main_conj} asserts existence without saying anything about the density of suitable traces. The problem remains open even with randomization: drawing random curves, as in the elliptic curve method~\cite{ecm}, heuristically yields a datum with $\Xi(\mathcal D)$ subexponential in $\log q$ in subexponential expected time, and a probabilistic polynomial-time construction with $\Xi(\mathcal D)\leq(\log q)^{O(1)}$ would also be a significant result.

\appendix

\section{Pseudo-code for the Mignotte--Schnorr algorithm}\label{sec:app_MS}

In this appendix, we recall the classical Mignotte--Schnorr algorithm, which uses the $\ell$-Sylow subgroups of $\F_q^\times$, for the primes $\ell$ dividing $q-1$, to find the roots of a split polynomial $f\in\F_q[X]$. This appendix contains no new results; rather, we reformulate the classical presentation of the Mignotte--Schnorr algorithm using the formalism developed throughout the paper. Classical references for these results include \cite{mignotte1988calcul} and \cite{von1987factoring}, while more modern formulations can be found in \cite[Sec.~2.2]{grenet2015randomized}.

We begin by defining a useful family of polynomials.
For $\beta\in\F_q^\times$, a prime number $\ell$, and $j\in\Z$, let $H_{\beta, \ell, j}(X)\in\F_q[X]$ be the polynomial whose roots are the elements $\zeta\in\F_q^\times$ such that $\ell$ does not divide $\order(\zeta\beta^{-j})$, i.e.
\begin{equation}\label{eq:H1}
H_{\beta,\ell,j}(X) := \prod_{\substack{\zeta\in\F_q^\times\\ \ell\,\nmid\,\order(\zeta\beta^{-j})}} (X-\zeta).
\end{equation}

These polynomials are used to isolate subsets of roots according to their projection onto the $\ell$-Sylow subgroup of $\F_q^\times$.

We now state the following useful lemma, which provides an explicit form for these polynomials. This description shows that $H_{\beta,\ell,j}(X)$ is a binomial whose roots form a coset of a subgroup of $\F_q^\times$. In particular, this makes it possible to compute modular reductions of $H_{\beta,\ell,j}(X)$ by first reducing a suitable power of $X$ using binary exponentiation.

\begin{lemma}\label{lem:property_Hl}
Let $\beta\in\F_q^\times$ and let $\ell$ be a prime number.
Write $\mu := \val_\ell(q-1)$ and $m := \val_\ell(\order(\beta))$.
For all $j\in\Z$,
\begin{equation}\label{eq:H2}
H_{\beta,\ell, j}(X) = X^{(q-1)/\ell^{\mu}} - \beta^{j(q-1)/\ell^{\mu}}.
\end{equation}
Moreover, for $j_1,j_2\in\Z$, we have $H_{\beta,\ell, j_1}(X) = H_{\beta,\ell,j_2}(X)$ if $j_1\equiv j_2 \bmod \ell^m$, and $H_{\beta,\ell,j_1}(X)$ is coprime with $H_{\beta,\ell, j_2}(X)$ otherwise.
\end{lemma}

\begin{proof}
Let $N_\ell:= (q-1)/\ell^{\mu}$ be the largest divisor of $q-1$ coprime with $\ell$. Since the right-hand sides of Equations~\eqref{eq:H1} and~\eqref{eq:H2} are monic separable polynomials, it suffices to show that they have the same roots in $\overline{\F_q}$.

An element $x\in\overline{\F_q}$ is a root of the right-hand side of~\eqref{eq:H2} if and only if
\[
(x\beta^{-j})^{N_\ell} = 1,
\]
that is, if and only if $x\beta^{-j}$ is an $N_\ell$-th root of unity. Since $N_\ell$ divides $q-1$, these roots of unity are exactly the elements of $\F_q^\times$ whose order divides $N_\ell$, i.e.\ is coprime with $\ell$. Hence $x$ is a root of the right-hand side of~\eqref{eq:H2} if and only if $x\in\F_q^\times$ and $\ell\nmid\order(x\beta^{-j})$.

Finally, let $x$ be a root of $H_{\beta,\ell, j_1}(X)$. By construction, $(x\beta^{-j_1})^{N_\ell} = 1$, hence
$$(x\beta^{-j_2})^{N_\ell} = (x\beta^{-j_1})^{N_\ell}\cdot \beta^{(j_1-j_2)N_\ell} = \beta^{N_\ell(j_1-j_2)}.$$
Since $\beta^{N_\ell}$ has order $\ell^m$, we have $\beta^{N_\ell(j_1-j_2)}= 1$ if and only if $j_1\equiv j_2 \bmod \ell^m$, and
we get that $H_{\beta,\ell, j_1}(X)$ and $H_{\beta,\ell,j_2}(X)$ have the same roots if $j_1\equiv j_2 \bmod \ell^m$, and they have no common root otherwise.
\end{proof}

\begin{corollary}
With the same notation as in Lemma~\ref{lem:property_Hl}, let $f(X)\in\F_q[X]$ be a polynomial. Then $\GCD(f(X), H_{\beta,\ell, j}(X))$ can be computed within $\Otilde(\log q \deg f)$ arithmetic operations in $\F_q$.
\end{corollary}

\begin{proof}
We first compute $H_{\beta,\ell, j}(X)\bmod f(X)$ using Lemma~\ref{lem:property_Hl}. To do so, we compute
\[
z(X) = X^{(q-1)/\ell^{\mu}} \bmod f(X)
\]
by binary exponentiation, which requires $\Otilde(\log q \deg f)$ operations. Then we obtain
\[
H_{\beta,\ell, j}(X)\bmod f(X) = z(X) - \beta^{j(q-1)/\ell^{\mu}}.
\]

Finally, we compute the GCD using the identity
\[
\GCD(f(X), H_{\beta,\ell, j}(X)) = \GCD(f(X), H_{\beta,\ell, j}(X)\bmod f(X)),
\]
which can be done in $\Otilde(\deg f)$ operations using fast GCD algorithms.
\end{proof}

The next proposition shows how the polynomials $H_{\beta, \ell, j}(X)$ can be used to deterministically separate the roots of a polynomial, once a primitive element of $\F_q^\times$ is known. This statement and its proof are similar to \cite[Lem.~3.1]{von1987factoring}.

\begin{proposition}
Let $\beta\in\F_q^\times$ be a primitive element. Write
\[
q-1 = \ell_1^{n_1}\cdots \ell_r^{n_r}
\]
for the prime factorization of $q-1$. Let $j_1\in\intrange{0}{\ell_1^{n_1}-1},\ldots, j_r\in \intrange{0}{\ell_r^{n_r}-1}$. Then there exists a unique $\alpha\in\F_q^\times$ such that, for all $i\in\intrange{1}{r}$, the order of $\alpha \beta^{-j_i}$ is not divisible by $\ell_i$, and
\[
(X-\alpha) = \GCD(H_{\beta, \ell_1, j_1}(X),\ldots, H_{\beta, \ell_r, j_r}(X)).
\]
\end{proposition}

\begin{proof}
This follows from the classical isomorphism
\[
\F_q^\times \simeq \Z/\ell_1^{n_1}\Z \times \cdots \times \Z/\ell_r^{n_r}\Z.
\]
Let $\alpha = \beta^\delta$, where $\delta\in\Z$ satisfies $\delta\equiv j_i \bmod \ell_i^{n_i}$ for all $i\in\intrange{1}{r}$. Such a $\delta$ exists and is unique modulo $q-1$ by the Chinese remainder theorem. Then $\alpha$ satisfies the desired property.
\end{proof}

\begin{algorithm}
  \caption{\CyclotomicPrimaryFacto{}: Deterministic Mignotte--Schnorr factorization with respect to the $\ell$-valuation of the orders of the roots.\label{algo:cyclotomicprimaryfacto}}
    \INPUT{
  \begin{itemize}[algo]
    \item A prime power $q$;
    \item A prime $\ell$;
    \item A nonnegative integer $m\leq \val_\ell(q-1)$\;
    \item A monic polynomial $f(X)\in\F_q[X]$ dividing $X^{(q-1)/\ell^{\val_\ell(q-1)-m}}-1$ (i.e.\ the $\ell$-valuations of the orders of all roots are at most $m$)\;
    \item An element $\beta\in\F_q^\times$ such that $\val_\ell(\order(\beta)) = m$.
  \end{itemize}
    }
    \OUTPUT{
  \begin{itemize}[algo]
        \item A finite subset $S\subset \intrange{0}{\ell^m-1}$\;
        \item A finite set of pairs $\{(j, f^{(j)}(X))\}_{j\in S}$ such that
      $f^{(j)}(X)\in\F_q[X]$ are monic nonconstant polynomials, $\prod_{j\in S} f^{(j)}(X) = f(X)$ and
      for all $j\in S$, 
      $$f^{(j)}(X) =
      \GCD(f(X), H_{\beta, \ell, j}(X)).$$
      \end{itemize}
    }
    \If{$f(X) = 1$}{
      Return $\{\}, \{\}$\;
    }
    \If{$m=0$}{
      $f^{(0)}(X) \gets f(X)$\;
      Return $\{0\}, \{(0, f^{(0)}(X))\}$\;
    }
    $S\gets \{\}$\;
    $g(X)\gets X^{(q-1) / \ell^{\val_\ell(q-1)-m+1}}\bmod f(X)$\label{line:cyclotomicprimaryfacto:modexp}\;
    $\hat\beta \gets \beta^{(q-1) / \ell^{\val_\ell(q-1)-m+1}}$\;
    \For{$i\in\intrange{0}{\ell-1}$}{
      $f_i(X)\gets \GCD(f(X), g(X)-\hat\beta^{i})$\label{line:cyclotomicprimaryfacto:gcd}\;
      $S_i, \{(j, f_i^{(j)})\}_{j\in S_i}\gets \CyclotomicPrimaryFacto(q, \ell, m-1, f_i(\beta^i\cdot X)\cdot \beta^{-i\cdot\deg(f_i)}, \beta^\ell)$\label{line:cyclotomicprimaryfacto:reccall}\;
      $S\gets S\cup \{\ell\cdot j + i : j\in S_i\}$\;
      \For{$j\in S_i$}{
        $f^{(\ell\cdot j+i)}(X) \gets f_i^{(j)}(\beta^{-i}\cdot X)\cdot\beta^{i\cdot\deg(f_i^{(j)})}$\label{line:cyclotomicprimaryfacto:subst}\;
      }
    }
    Return $S, \{(j, f^{(j)}(X))\}_{j\in S}$.
\end{algorithm}

We provide a proof of termination and correctness of the algorithm, as similar techniques are used in the proof of Proposition~\ref{prop:twoadicfactowitheval}.

\begin{proposition}
  Algorithm~\ref{algo:cyclotomicprimaryfacto} (\CyclotomicPrimaryFacto{}) terminates and is correct.
\end{proposition}

\begin{proof}
Throughout the proof, we use the notation $\mu := \val_\ell(q-1)$.

\smallskip

The termination is proved by induction on $m$. For $m=0$, the algorithm clearly terminates. For $m > 0$, each recursive call decreases the value of $m$, and hence the algorithm eventually terminates.

\smallskip

The correctness is also proved by induction on $m$. For $m=0$, the assumption that $f(X)$ divides $X^{(q-1)/\ell^\mu}-1$ implies that the orders of all roots of $f(X)$ are not divisible by $\ell$. Hence $f(X)=f^{(0)}(X)$, and the algorithm is correct in this case.

Let $m \geq 1$, and assume that the algorithm is correct at level $m-1$. We prove correctness at level $m$. To do so, we must show that the inputs of the recursive calls satisfy the required assumptions, and that the output satisfies the desired properties.

\smallskip

First, we prove that for all $i\in\intrange{0}{\ell-1}$, the inputs of the recursive call at Line~\ref{line:cyclotomicprimaryfacto:reccall} satisfy the required assumptions at level $m-1$. Since $\val_\ell(\order(\beta)) = m\geq 1$, we have $\val_\ell(\order(\beta^\ell)) = m-1$. The fact that $f_i(\beta^i X)$ divides $X^{(q-1)/\ell^{\mu-m+1}}-1$ follows from the relation $f_i(X)\mid g(X)-\hat\beta^i$, which implies that $f_i(X)$ divides $(\beta^{-i}X)^{(q-1)/\ell^{\mu-m+1}}-1$.

\smallskip

Next, we prove that the output is correct using the induction hypothesis. Since $\hat\beta$ has order $\ell$, we first observe that
\[
X^{(q-1)/\ell^{\mu-m}}-1 = c\prod_{i\in\intrange{0}{\ell-1}} \left((\beta^{-i}X)^{(q-1)/\ell^{\mu-m+1}} - 1\right),
\]
where $c := \hat\beta^{\ell(\ell-1)/2}$ is a nonzero constant (equal to $-1$ if $\ell=2$, and to $1$ otherwise). Since the factors of this product are pairwise coprime, this implies that $\prod_{i\in\intrange{0}{\ell-1}} f_i(X) = f(X)$.

Consequently,
\begin{align*}
\prod_{j\in S} f^{(j)}(X)
&= \prod_{i\in\intrange{0}{\ell-1}}\ \prod_{j\in S_i}
f_i^{(j)}(\beta^{-i} X)\cdot \beta^{i\deg(f_i^{(j)})} \\
&= \prod_{i\in\intrange{0}{\ell-1}} f_i(X) = f(X).
\end{align*}

Let $j\in S$, and let $i\in\intrange{0}{\ell-1}$ be such that $j\equiv i \bmod \ell$, so that $\lfloor j/\ell\rfloor\ell = j-i$. By the induction hypothesis, applied to the recursive call on $f_i(\beta^i X)$ with $\beta^\ell$ in place of $\beta$, we have
\[
f_i^{(\lfloor j/\ell\rfloor)}(X)
= \GCD\bigl(f_i(\beta^i X), H_{\beta^\ell,\ell,\lfloor j/\ell\rfloor}(X)\bigr)
= \GCD\bigl(f_i(\beta^i X), H_{\beta,\ell,j-i}(X)\bigr),
\]
since $(\beta^\ell)^{-\lfloor j/\ell\rfloor} = \beta^{i-j}$. Moreover, the change of variable $\zeta\mapsto\beta^i\zeta$ in~\eqref{eq:H1} shows that $H_{\beta,\ell,j-i}(\beta^{-i}X)$ equals $H_{\beta,\ell,j}(X)$ up to a nonzero multiplicative constant. Therefore, up to such a constant,
\[
f^{(j)}(X)
= f_i^{(\lfloor j/\ell\rfloor)}(\beta^{-i} X)
= \GCD\bigl(f_i(X), H_{\beta,\ell,j}(X)\bigr).
\]

We conclude by noticing that $f(X) = f_0(X)\cdots f_{\ell-1}(X)$, and that $i\not\equiv j\bmod \ell$ implies that $f_i(X)$ is coprime to $H_{\beta, \ell, j}(X)$. Therefore,
\[
f^{(j)}(X)=\GCD(f_i(X), H_{\beta,\ell,j}(X)) = \GCD(f(X), H_{\beta,\ell,j}(X)).
\]
\end{proof}

\begin{proposition}\label{prop:cyclotomicprimaryfacto}\cite{mignotte1988calcul}
  Algorithm~\ref{algo:cyclotomicprimaryfacto} (\CyclotomicPrimaryFacto{}) performs at most $$\Otilde\bigl(\deg(f) m (\ell +\log q)\bigr)$$ arithmetic operations in $\F_q$.
\end{proposition}

\begin{proof}
We refer to \cite[Proof of Thm.~1]{mignotte1988calcul} or \cite[Proof of Lem.~3]{grenet2015randomized}.
\end{proof}

\bibliographystyle{alpha}
\bibliography{biblio}

\end{document}